\documentclass[12pt,preprint]{aastex}

\newcommand{\msol}{\hbox{\kern 0.20em $M_\odot$}}
\newcommand{\lsol}{\hbox{\kern 0.20em $L_\odot$}}
\newcommand{\rsol}{\hbox{\kern 0.20em $R_\odot$}}
\newcommand{\sr}{\hbox{\kern 0.20em sr}}
\newcommand{\srmu}{\hbox{\kern 0.20em sr$^{-1}$}}
\newcommand{\g}{\hbox{\kern 0.20em g}} 
\newcommand{\gmu}{\hbox{\kern 0.20em g$^{-1}$}}
\newcommand{\kg}{\hbox{\kern 0.20em kg}}
\newcommand{\pc}{\hbox{\kern 0.20em pc}} 
\newcommand{\mum}{\hbox{\kern 0.20em $\mu$m}}
\newcommand{\mumd}{\hbox{\kern 0.20em $\mu$m$^{-2}$}}
\newcommand{\cm}{\hbox{\kern 0.20em cm}}
\newcommand{\m}{\hbox{\kern 0.20em m}} 
\newcommand{\km}{\hbox{\kern 0.20em km}}
\newcommand{\nm}{\hbox{\kern 0.20em nm}}

\newcommand{\s}{\hbox{\kern 0.20em s}} 
\newcommand{\h}{\hbox{\kern 0.20em h}}
\newcommand{\smu}{\hbox{\kern 0.20em s$^{-1}$}}
\newcommand{\smd}{\hbox{\kern 0.20em s$^{-2}$}}
\newcommand{\an}{\hbox{\kern 0.20em an}} 
\newcommand{\anmu}{\hbox{\kern 0.20em an$^{-1}$}}
\newcommand{\yr}{\hbox{\kern 0.20em yr}} 
\newcommand{\yrmu}{\hbox{\kern 0.20em yr$^{-1}$}}
\newcommand{\Myr}{\hbox{\kern 0.20em Myr}} 
\newcommand{\Mymu}{\hbox{\kern 0.20em Myr$^{-1}$}}
\newcommand{\K}{\hbox{\kern 0.20em K}}  
\newcommand{\pcmu}{\hbox{\kern 0.20em pc$^{-1}$}}
\newcommand{\pcmd}{\hbox{\kern 0.20em pc$^{-2}$}}
\newcommand{\pcmt}{\hbox{\kern 0.20em pc$^{-3}$}}
\newcommand{\kms}{\hbox{\kern 0.20em km\kern 0.20em s$^{-1}$}}
\newcommand{\kmpd}{\hbox{\kern 0.20em km$^{2}$}}
\newcommand{\kpc}{\hbox{\kern 0.20em kpc}}
\newcommand{\cms}{\hbox{\kern 0.20em cm\kern 0.20em s$^{-1}$}}
\newcommand{\erg}{\hbox{\kern 0.20em erg}}
\newcommand{\ergs}{\hbox{\kern 0.20em erg}}
\newcommand{\cmpd}{\hbox{\kern 0.20em cm$^2$}}
\newcommand{\cmmd}{\hbox{\kern 0.20em cm$^{-2}$}}
\newcommand{\cmms}{\hbox{\kern 0.20em cm$^{-6}$}}
\newcommand{\cmpt}{\hbox{\kern 0.20em cm$^3$}}
\newcommand{\cmmt}{\hbox{\kern 0.20em cm$^{-3}$}}
\newcommand{\mpd}{\hbox{\kern 0.20em m$^2$}}
\newcommand{\mmd}{\hbox{\kern 0.20em m$^{-2}$}}
\newcommand{\mpt}{\hbox{\kern 0.20em m$^3$}}
\newcommand{\mmt}{\hbox{\kern 0.20em m$^{-3}$}}
\newcommand{\mujy}{\hbox{\kern 0.20em $\mu$Jy}}
\newcommand{\mjy}{\hbox{\kern 0.20em mJy}}
\newcommand{\Mj}{\hbox{\kern 0.20em MJy}}
\newcommand{\jy}{\hbox{\kern 0.20em Jy}}
\newcommand{\ghz}{\hbox{\kern 0.20em GHz}}
\newcommand{\G}{\hbox{\kern 0.20em G}}
\newcommand{\muG}{\hbox{\kern 0.20em $\mu$G}}
\newcommand{\twco}{\hbox{${}^{12}$CO}}

\newcommand{\thco}{\hbox{${}^{13}$CO}}

\newcommand{\ceio}{\hbox{C${}^{18}$O}}

\newcommand{\cs}{\hbox{CS}}

\newcommand{\htwo}{\hbox{H${}_2$}}

\newcommand{\hcop}{\hbox{HCO$^{+}$}}
\newcommand{\htcop}{\hbox{H$^{13}$CO$^{+}$}}
\newcommand{\sulfp}{\hbox{[S{\small II}]}}
\newcommand{\sulfpp}{\hbox{[S{\small III}]}}
\newcommand{\Hp}{\hbox{H{\small II}}}
\newcommand{\Cp}{\hbox{[C{\small II}]}}
\newcommand{\Opp}{\hbox{[O{\small III}]}}
\newcommand{\Oi}{\hbox{[O{\small I}]}}
\newcommand{\Np}{\hbox{[N{\small II}]}}
\newcommand{\Nep}{\hbox{[Ne{\small II}]}}
\newcommand{\Nepp}{\hbox{[Ne{\small III}]}}
\newcommand{\Npp}{\hbox{[N{\small III}]}}
\newcommand{\Sip}{\hbox{[Si{\small II}]}}
\newcommand{\jonetozero}{\hbox{$J=1\rightarrow 0$}}
\newcommand{\jtwotoone}{\hbox{$J=2\rightarrow 1$}} 
\newcommand{\jthreetotwo}{\hbox{$J=3\rightarrow 2$}}
\newcommand{\jfivetofour}{\hbox{$J=5\rightarrow 4$}}

\shortauthors{Lefloch et al.}

\shorttitle{Photoionization of a Star-Forming Core}
\begin{document}

\title{The Photoionization of a Star-Forming Core \\ 
in the Trifid Nebula}
\author{
B. Lefloch\altaffilmark{1}, 
J. Cernicharo\altaffilmark{2},
\and
L. F. Rodr\'{\i}guez\altaffilmark{3}, 
M.A. Miville-Desch\^enes\altaffilmark{4},
D.~Cesarsky\altaffilmark{5}, 
A.~Heras.\altaffilmark{6}
}

\altaffiltext{1}{Laboratoire d'Astrophysique, 
Observatoire de Grenoble, BP 53, F-38041, Grenoble Cedex 9, France
\email{lefloch@obs.ujf-grenoble.fr}
}

\altaffiltext{2}{Consejo Superior de Investigaciones Cient\'{\i}ficas,
Instituto de Estructura de la Materia, Serrano 121, E-28006 Madrid}

\altaffiltext{3}{Instituto de Astronom\'{\i}a, UNAM, 
Campus Morelia, A.P. 3-72, Morelia,   Mich. 58089, M\'exico}
\altaffiltext{4}{ENS, LRA, 24 rue Lhomond, Paris, France}
\altaffiltext{5}{Max-Planck Institut f\"ur Extra-Terrestrische Physik, 
85741 Garching, Germany}
\altaffiltext{6}{ESTEC/SCI-SAF, P.O. Box 209, NL-2200 AG Noordwijk The 
Netherlands}

\begin{abstract}

We have carried out a comprehensive multiwavelength study of
Bright-Rimmed Globule TC2 in the Trifid Nebula, using the IRAM~30m
telescope, the VLA centimeter
array and the Infrared Space Observatory (ISO).
TC2 is one of the very few globules to exhibit signs of active ongoing
star formation while being photoevaporated by the Ly-c flux of the exciting
star of the nebula ($\sim 10^{10}\cmmd\smu$).
The globule consists of a cold dense core of mass $27\msol$ surrounded by
a lower density envelope of molecular gas. The impinging Ly-c photons
induce the propagation of an ionization front into the globule.
The evaporation of the ionized gas forms a thin layer of density
$n_e= 1-2\times 10^3\cmmt$ around the globule,
which could be mapped with the VLA.
The globule is illuminated mainly on its rear side,
by a FUV field of intensity $\rm G_0\simeq 1000$.
It creates a Photon-Dominated Region (PDR) below the surface,
which was mapped and characterized with the ISOCAM Circular Variable Filter
and the Short Wavelength Spectrometer (SWS) onboard the Infrared Space
Observatory.
The physical conditions derived from the analysis of the far-infrared
lines $\Oi$~$\rm 63\mu m$, $\rm 145\mu m$ and $\Cp$~$\rm 158\mu m$,
and the continuum
emission are in good agreement with some recent PDR models.
The emission of the PAHs band at 6.2, 7.7, 8.6 and $\rm 11.3\mu m$ is
detected over the whole globule. The relative intensity variations 
observed across the globule, in the PDR and the photoionized envelope, 
are consistent with the changes in the ionization fraction.
In the head of TC2, we find a second  kinematic component
which is the signature of the radiatively-driven collapse undergone by the 
globule. This component indicates that the PDR propagates at low velocity
inside the body of TC2. 
The molecular emission suggests that the star formation process
was probably
initiated a few $10^5\yr$ ago, in the large burst which led to the
formation of the nebula. 
The globule has already evaporated half the mass of its envelope. 
However, the ionization timescale of the globule is long enough 
($\sim 2\Myr$) to let the newly born object(s) reach smoothly the ultimate 
stages of protostellar evolution. 
The impact of photoionization on the star formation process
appears limited. 
\end{abstract}

\keywords{ ISM: globules --- dust, extinction --- HII regions --- 
ISM: individual (Trifid) --- ISM: jets and outflows --- 
stars: formation}

\section{Introduction}

It is well established that the bright-rimmed globules found in 
\Hp\ regions are often sites of star formation. Reipurth (1983) first 
showed that these objects do form stars and subsequent work based on 
IRAS data by Sugitani et al. (1991) confirmed that they are indeed 
active ``stellar factories'' which produce intermediate-mass (Herbig AeBe) 
stars. 
These condensations are local clumps which emerge from the expanding 
nebula or form from the fragmentation of the dense molecular layer 
surrounding the ionized gas. The various theoretical works 
(Bertoldi, 1989; Bertold \& McKee, 1990; Lefloch \& Lazareff, 1994) and 
and the numerous observational studies on bright-rimmed globules 
(see e.g. Cernicharo et al. 1992; Lefloch \& Lazareff, 1995)
have enabled to draw the following evolutionary picture, summarized in 
Fig.~1. 

As a globule of neutral gas is exposed to the ionizing field of the 
exciting star(s) of the nebula (region I in Fig.~1), an ionization front 
(IF) forms at the surface of the condensation. For standard ionization 
conditions, the pressure of the surface ionized gas is much higher 
than in the nebular gas and in the molecular globule. As a consequence, 
the photoionized gas expands into the 
H{\small II} region, inducing the formation of a photoionized envelope 
around the globule. It is the ionization front and the photoionized envelope 
which are detected in the optical as a bright rim (region II). 
The incident FUV field drives the formation of a Photon-Dominated Region
(hereafter PDR, region III) while a shock front, driven by the surface 
overpressure, propagates
towards the dense molecular core (region IV). Observational 
evidence of this mechanism, also called Radiatively-Driven Implosion (or RDI), 
has been reported in a few objects by Cernicharo et al. (1992) and  
Lefloch \& Lazareff (1995). Progressively, a dense core forms behind the 
surface, and the globule adjusts its internal structure to balance the 
pressure of the ionized gas, while the bulk of its mass is photoevaporated.
Eventually, the globule reaches a quasi 
steady-state, the ``cometary phase'', in which the shock front has 
disappeared. The globule now consists of a small dense ``head''
prolonged by a long tail of diffuse gas. 

It has long been suggested that the shock front inside the globule
could trigger the star formation inside the bright-rimmed globules 
(see e.g. Reipurth 1983; Lefloch et al. 1997) . These objects appear 
therefore as ideal laboratories to test the scenarios of star 
formation triggered by an external compression wave. Most of the 
studies led until now were focused onto the molecular core of bright-rimmed 
globules (see e.g. Lefloch et al. 1997). Therefore, the  
physical conditions reigning in the PDR and in the shocked molecular gas 
are not well characterized and the impact of photoionization on the 
gravitational collapse is therefore difficult to evaluate. 
Moreover, all the bright-rimmed condensations studied until now are 
found in relatively old \Hp\ regions, with ages of a few Myr. 
Because the condensations are usually found at rather large distances from 
the ionizing stars, they experience a reduced UV field and it is difficult to 
discriminate between a star formation induced by Radiatively-Driven
Implosion and 
a spontaneously evolving globule which has already started to form stars 
by the time it is hit by the ionization front.

This is why we have started a systematic multiwavelength study 
of a young H{\small II} region~: the Trifid nebula. It appears as a small 
dusty nebula of $10\arcmin$ diameter
at an heliocentric distance of $1.68\kpc$ (Lynds et al. 1985), with a 
dynamical age of $0.3-0.4\Myr$ . The nebula is excited by the O star 
HD~164492A. In a preliminary work (Cernicharo et al. 1998, CL98), we
reported on the mapping of the thermal dust emission of the nebula at
millimeter wavelengths. This mapping revealed the presence of several
protostellar cores (dubbed TC1 to TC4) in the shell of dense molecular gas
surrounding the \Hp\ region. It was not clear however if the birth of the 
protostars was triggered or not. For this reason, we undertook a more
detailed analysis of the protostellar cores. 
In a subsequent paper (Lefloch \& Cernicharo, 2000), we reported on the
continuum and molecular line emission around the two most massive
protostellar cores~: TC3-TC4. Their masses are high, between 60 and
$90\msol$. They harbour a Class I source and one of the few high-mass
Class~0 candidates known until now. 
Comparison of their properties with the models of Elmegreen \& Lada (1977) 
and Whitworth et al. (1994)  allowed to conclude that the formation of TC4
had probably been triggered in the fragmentation of the dense shell 
surrounding the ionized gas. 
The molecular properties of TC3 and TC4 are similar to those of the 
protostellar cores discovered in Orion, though at an earlier, ``pre-Orion'', 
evolutionary stage.

We report here on the TC2, protostellar core, which is associated with a 
bright-rimmed globule on the Southern border of the Trifid. TC2 appears 
to be in a more advanced 
stage of photoionization than TC3-4~: unlike TC3-4 which are still
embedded cores, TC2 has already emerged from the diffuse molecular gas
layers  and it exhibits the optical bright rim and the cometary shape
typical of photoionized globules. As discussed in this work, TC2 is exposed 
to a rather strong ionizing field, which drives a shock into the condensation, 
in agreement with the evolutionary scheme presented above. 
Like TC3 and TC4, TC2 displays signs of protostellar activity. 
CL98 reported the presence of the Herbig-Haro jet HH399 coming out of the 
head of the globule and propagating into the ionized nebula. 
It is the best example
of globules undergoing at the same time strong photoevaporation and active 
star formation. It offers a good opportunity to study the quantify the 
relation both phenomenons hold to each other, and better constrain the role
that Radiatively-Driven Implosion could play in the star formation process.
In this purpose,
we have led a detailed study of the structure and the physical conditions 
of the globule (density, temperature, velocity). Because TC2 
lies almost in the plane of the sky, it provides also a good opportunity to 
study the structure of a typical low-density photon-dominated region, and 
to confront its 
properties against the existing models. This work provides the first 
comprehensive study of the whole gas structure of a bright-rimmed globule, 
from the ionized surface layers to the cold dense molecular core.

The paper is organized as follows.
We first derive the structure of the globule~: 
The HII region (Sect.~3);
the bright rim and the photo-evaporated envelope (Sect.~4); the PDR
(Sect.~5); the dust continuum emission (Sect.~6)
 and the molecular core (Sect.~7). We then discuss the observational
evidences of Radiatively-Driven Implosion in TC2 and the implications on the 
past history of the globule (Sect.~8). In the following section, 
we first study the star forming conditions in TC2 and attempt to 
characterize the protostellar source and the outflowing material, 
before studying the impact of photo-ionization 
both on the protostar and the evolution of the globule (Sect.~9). 
The conclusions are presented in Sect.~10. 

\section{The Observations}

\subsection{Observational Approach}

The determination of the physical conditions in the various regions in 
TC2 requires a compared study of the spectral line and continuum 
emission mapped at various wavelengths.
The photoionized envelope at the surface of TC2 was characterized  
from observing the radio free-free emission with the Very Large Array (VLA), 
following the same approach as Lefloch et al. (1997). Additional constraints
on the envelope and the surrounding nebular gas in the Trifid were brought 
by the fine-structure atomic and ionic lines detected in the mid-IR and 
FIR with the Infrared Space Observatory (ISO; Kessler et al. 1996).
The conditions in the molecular core were derived from the emission of the 
cold dust and of various molecular tracers at millimeter wavelengths with the 
IRAM 30m telescope. 
The physical conditions in the PDR were determined mainly from the 
observations of the FIR $\Oi$ lines at $\rm 63\mu m$ and 
$\rm 145\mu m$ and the 
$\Cp$ line at $\rm 158\mu m$. The dust properties 
were derived from the observation of the continuum emission between 5 and 
$\rm 197\mu m$ with the instruments onboard ISO. 
In all maps presented here the coordinates of
all the maps are given in arcsec offsets with respect to the position of 
HD~164492A~: $\alpha(2000)=$ $18^{\rm h} 02^{\m} 23.55^{\s}$, 
$\delta(2000)= -23^{\circ} 01\arcmin 51\arcsec$.
The observational data set is summarized in Table~1; 
the data described below are shown in Figures 2 to 9 and 11 to 16. 

\subsection{The VLA data}

The Very Large Array observations at 3.6~cm were made with the array
in its highest angular resolution A configuration in 1998 March 13.
We used 1328+307 as absolute amplitude calibrator and 1748-253 as the
phase calibrator. A bootstrapped flux density of 0.271$\pm$0.001 Jy was 
obtained for 1748-253.
The observations were made in both circular polarizations with an
effective bandwidth of 100 MHz.  The data were edited and calibrated
following the standard VLA procedures and using the software package
AIPS. The final map at 3.6cm was made with the AIPS task IMAGR, the ROBUST 
parameter set to 5, and a Gaussian taper of 100 k$\lambda$ to
the (u,v) data to lower the angular resolution. The field of view is 
approximately $5\arcmin$; the synthesized beam has a 
size of 
$1{\rlap.}{^{\prime \prime}}7 \times 1{\rlap.}{^{\prime \prime}}4$ (HPFW) and
makes a position angle (P.A.) of $17^\circ$.

\subsection{The Millimeter Continuum Observations}

Observations of the 1.25mm continuum emission in the Trifid were carried out
in March 1996 and March 1997 at the IRAM 30m-telescope (Pico Veleta, Spain)
using the MPIfR 19-channel bolometer array. The resolution of the telescope
is $11\arcsec$. The final map was obtained by combining several individual
fields centered on the brightest condensations of the nebula. Each field was
scanned in the horizontal direction by moving the telescope at a speed of
$4\arcsec$ per sec; subsequent scans  are displaced by $4\arcsec$. We used
a chopping secondary at 2Hz with a throw of 30 to $60\arcsec$ depending on
the structure of the region to be mapped.  Calibration was checked against
Mars. The weather conditions were good and rather stable during the two
observing sessions. The opacity was monitored every hour on average and
we found typical zenith opacities between 0.1 and 0.35. Pointing was checked
every hour as the source transits at low elevation at Pico Veleta and
corrections were always found to be lower than $3\arcsec$. The final rms 
is $8\mjy/11\arcsec$~beam. Hence, it is sensitive enough to detect at the
$3\sigma$ level protostellar condensations of $1\msol$ at 20~K in one 
$11\arcsec$ telescope beam ($0.09\pc$ at the distance of the Trifid). 
In order to outline the weak extended dust components
in the nebula, we degraded the angular resolution of our map down to 
$15\arcsec$ (0.13\pc\ at the distance of the Trifid) by convolving the 
emission with a gaussian of 11\arcsec\ HPFW. The resulting rms in the map 
is $5\mjy\, \rm beam^{-1}$. All the results quoted in 
this paper are based on the non-degraded map with 0.09\pc\ resolution. 

\subsection{The Molecular Line Observations}

We observed in July 1996 and July 1997 the Trifid nebula in the
millimeter lines of SiO, \hcop, \htcop\  and CS, 
with the IRAM 30m-telescope. The lines were observed with a spatial 
sampling of $15\arcsec$~: the 
data are almost Nyquist-sampled at 3mm. We used an autocorrelator
as spectrometer, which provided a velocity resolution of $\approx 0.2\kms$ in 
all three bands. The rejections of the receivers were always higher than 
10dB, and checked against W51D (Mauersberger et al. 1989). 
The Trifid was mapped at full sampling in the CO $\jtwotoone$,
$\jonetozero$, $\thco$ $\jonetozero$ and $\ceio$ 
$\jonetozero$ lines with the IRAM 30m telescope in July 1996. 
The autocorrelator provided a kinematic resolution of $0.2\kms$ for
all the transitions. Additionnal observations of the CO \jthreetotwo\ 
transition were carried out at the CSO during various observing runs
between 1998 and 1999. The receiver was connected to an AOS which provided 
a kinematic resolution of $0.4\kms$. All the observed lines, their frequency, 
the telescope beamwidth and the main-beam efficiencies are summarized 
in Table~2.

\subsection{The Mid-Infrared Observations}

We observed the Trifid nebula with ISOCAM (Cesarsky et al. 1996) onboard ISO. 
An image of the whole nebula was obtained using the broad band LW10 filter 
($\lambda= 11.5\mu m$, $\Delta\lambda= 7\mu m$) with a pixel size of 
3\arcsec. The mid-infrared emission around TC2 was observed in the 
Circular Variable Filter mode (CVF)
between 5 and $17\mu m$. The pixel size is 
$1.5\arcsec\times 1.5\arcsec$ and the spectral resolution is 40. The 
$32\times 32$ pixel detector covered a total field of view of 
$48\arcsec\times 48\arcsec$ 
centered on the globule at $\alpha(2000)=$ $18^{\rm h} 02^{\m} 28.7^{\s}$, 
$\delta(2000)= -23^{\circ} 03\arcmin 51\arcsec$.
The size of the Point Spread Function varies between $\approx 1.5\arcsec$ 
at $5\mu m$ and $\approx 6\arcsec$ at $17\mu m$.
The data were reduced with the SLICE software and
following the method of Miville-Deschenes et al. (2000).

The emission between 2 and $45\mu m$ was observed with the
SWS spectrometer onboard ISO (de Graauw et al., 1996) in the SW01 mode 
($2.4 - 45\mu m$ grating scan)
The spectral resolution was 300. A spectrum was taken right on the
head of TC2; two additional spectra were taken at a position shifted by 
+20\arcsec\ and - 20\arcsec\ in declination, in order to 
measure the emission towards the bright rim and the H{\small II} region and
towards the main body of the globule respectively. The positions are marked
in Fig.~4. 
We note that there is some overlap between the three beams. 
In particular the beam centered $20\arcsec$ North of the globule 
encompasses the northern part of the bright rim. 
The SWS data consists of an ``up'' scan, towards decreasing wavelengths,
and a ``down'' scan towards increasing wavelengths.
The lines identified
and their flux (obtained after averaging the ``up'' and ``down'' scans) are
given in Table~3. 
The calibration accuracy varies from $5\%$ at $2.5\mu m$ to $30\%$ at $45\mu m$
(SWS ISO Handbook, Leech et al. 2001). 
Because of the high noise, the statistical errors in the line
fluxes are {\em a priori } not negligible in front of the systematic errors
in the flux calibration. Therefore, we have compared the emission in
the ``up'' and ``down'' scans in order to estimate the 
actual uncertainty in the line fluxes. For all the lines detected, {\em
but the $\Nepp$ line at $15.55\mu m$}, a very good agreement was found 
between both measurements. On the contrary, very large variations, up to a
factor of 4,  were observed in the flux of the $\Nepp$ line. 

The emission between 45 and $\rm 197\mu m$ was observed  
towards the globule with the LWS spectrometer onboard ISO (Clegg et al. 
1996). 
The size of the LWS beam is known to vary between $66\arcsec$ and $86\arcsec$ 
HPFW depending on the detector's band. This effect was taken into account to 
estimate the line fluxes~: the detector beam sizes were taken from the 
ISO Handbook for the LWS spectrometer (Gry et al., 2001). 
The large-scale emission of the nebula 
was estimated by observing a nearby reference position $90\arcsec$ away 
of TC2, at position $\alpha(2000)= 18^h02^m35.1^s$, 
$\delta(2000)= -23^{\circ}03'51.8\arcsec$ (see Fig.~2). The data was 
reduced using the Off The Line package (OLP) version 7 and with the ISAP 
package. The fluxes measured are given 
following the standard flux calibration, assuming a 
point-like source. The uncertainty in the absolute flux calibration quoted 
for LWS is $10-15\%$. However, as discussed in the text, part of the emission
in the LWS beams arises from the nebular gas, i.e. very extended
emission which fills the beam. Therefore, we adopt in this case a somewhat
more conservative number of $20\%$. 
The identified lines and the fluxes are listed in 
Table~\ref{flux_lws}. 
The spectrum of the globule was obtained after subtracting
the emission of the reference position (OFF source) to the emission 
measured ON source.

\section{The HII region : zone I}

We proceed first with the analysis of the $\Hp$ region as TC2 is embedded
in the ionized gas and an estimation of the physical conditions of
the ionized bright rim (see next section) requires a knowledge of
the contribution to the atomic fine structure lines from the HII region.

\subsection{Emission from the nebular gas}

We show in Fig.~6 the various atomic and fine-structure atomic lines
detected with the LWS and SWS spectrometers. 
Many of the identified lines exhibit weak variations between the On and Off
positions. 

The best example is provided by $\Cp$ at $\rm 158\mu m$. On the other
hand, the largest variations are observed for the $\Oi$ lines at 63.3 and 
$\rm 145.5\mu m$. In particular, almost no emission at all is detected
in the $\rm 145\mu m$ line at the reference position

The $\Opp$ $52, 88\mu m$ and $\Npp$ $\rm 57\mu m$ lines are useful tools 
to diagnose the electron density in the ionized gas. 
The  ($\Opp$ $52/88$) and ($\Opp ~52$/$\Npp ~57$) ratios
have the advantage of being rather insensitive to the temperature, and mainly 
trace the electron density $\rm n_e$ in the ionized gas.
Like for the SWS data, the situation is made complicated by the fact 
that the LWS beam encompasses regions with very different
physical conditions. In order to gain more
insight on the background emission from the $\Hp$ region and the PDR
associated with the parent molecular cloud, we first consider the 
reference position.

The LWS flux scale is based on a point source calibration. In the case of 
extended source emission, some correcting factors have to be applied 
to derive the correct fluxes (see paragraph 4.9.3 in the ISO
Handbook for the LWS spectrometer, Gry et al. 2001). 
Since the emission at the reference position comes from the $\Hp$ region 
and fills the whole LWS beam, the fluxes have been calibrated  
according to the ISO manual prescription. We then 
obtain $(\Opp ~52/88)= 0.69$. This corresponds to a mean
density $\rm n_e\simeq 50\cmmt$, which is consistent with the previous
determination from $\sulfpp$. In this density range, the emissivity of the 
$\rm 52\mu m$  line is ~: 
$\epsilon_{52}= 8.0\times 10^{-23}\erg\cmpt\smu\srmu$ 
(Dinerstein et al., 1985).
If we assume a standard abundance $\rm \Opp/[H^{+}]= 2\times 10^{-4}$, 
we derive an estimate of the emissivity 
$\int n_e^2 dl= 2.6\times 10^3 \rm cm^{-6} \pc$ and the size of the 
emitting region~: $l\simeq 1\pc$. 

In the density range measured in the \Hp\ region, the emissivity ratio
($\Npp ~57/ \Opp ~52$) is predicted to be close to 1.6 
(see e.g. Lester et al. 1987), whereas the measured line ratio is only 
0.88. Since both lines trace physically similar regions 
(their critical densities are 
$1.88\times 10^3\cmmt$ and $3.25\times 10^3\cmmt$ respectively)
this discrepancy mainly reflects the different relative elemental 
abundances. We obtain an elemental abundance ratio 
($\Npp / \Opp)\simeq 0.5$. 
This ratio agrees with other determinations in a sample of \Hp\ regions 
by Lester et al. (1983) who obtained 0.25-0.43.

In the SWS spectra, the line $\Nep$ $\rm 12.7\mu m$ is
detected at the three observed positions~: towards the head, 
the bright rim, and the main
body of the globule. The fluxes exhibit only 
weak variations between the three positions.  This suggests 
that the contribution of the $\Hp$ region, which fills
the SWS beams, dominates the emission from the globule. 
The SWS fluxes
of the  $\Nepp$ $15.5\mu m$ line suffer large statistical errors which
casts some
doubt on the variations measured with the SWS~: whereas the fluxes in the
Northen and Central positions are similar (better than 30\%), there seems
to be hardly any emission detected in the Southern position.

Similarly to the $\Nep$ line, the $\sulfpp$ lines at 18.7 and $33.7\mu m$
exhibit only weak variations between the three positions observed with 
SWS. In order to determine the origin of the $\sulfpp$ emission, 
we first estimate the electron density from the 
$\sulfpp$ $\rm 18/33\mu m$ ratio. In the case of extended sources, a 
correcting factor has to be applied to the SWS flux obtained with the 
reduction pipeline in order to derive the correct source flux (Salama 2000). 
It is difficult to estimate the interstellar absorption on the line of
sight. Since the
emission arises from the region close to the {\em optical} bright rim, 
we believe the interstellar extinction should be rather low and we will 
neglect the latter in what follows.  
We estimate an emissivity ratio 
$\epsilon_{18.7}/\epsilon_{33.7}\simeq 0.7$ for the Northern spectrum. 
This ratio allows only to set an upper limit on the electron density
(Rubin 1989)~: $\rm n_e \leq 200\cmmt$. 
If the contribution of the bright rim were dominant, 
one would obtain a much larger emissivity ratio, of 
$\sim 2$, as expected for the electron densities obtained with the VLA 
($\sim 10^3\cmmt$). 
This agrees with previous determinations 
of the {\em nebular} gas density by Lynds et al. (1985)  and Rosado 
et al. (1999) from the optical [S{\small II}] $\rm 6717,6731 \AA$ lines
and with our determination from the atomic fine structure lines detected
with the LWS (see above).
 We conclude that the main contribution to the 
$\sulfpp$ flux arises from the $\Hp$ region. 

\subsection{The Effective Stellar Temperature of HD~164492A}

Simpson et al. (1995) showed that it is possible to use the 
ionization fraction ratios of heavy elements such as S, O or N in order to 
constrain the effective stellar temperature $\rm T_{eff}$ by comparison with 
models of $\Hp$ regions.  In particular Rubin (1985) calculated the
integrated fluxes of numerous ionic lines of such heavy elements for a 
wide range of conditions including the neutral gas density and the 
effective stellar temperature.  These models predict that the 
ionization fraction ratio $\rm <O^{2+}/O>/<S^{2+}/S>$ 
spans 3 orders of magnitude when the effective temperature of the exciting
star ranges from 3 to  $\rm 4\times 10^4~K$ 
(see e.g. Fig.~8 in Simpson et al. 1995). 

The method to derive the ionization fraction ratio is fully discussed in 
Simpson et al. (1995). 
We first estimate the ionic ratio $\rm O^{2+}/S^{2+}$ from the 
$\Opp$ lines at 52 and $\rm 88\mu m$ and the $\sulfpp$ $19\mu m$ line.
The emissivities for $\Opp$ and $\sulfpp$ are taken from 
Dinerstein et al. (1985) and Rubin et al. (1994) respectively. 
We adopted $\rm O/S= 47$ as elemental abundance, which is the value found 
in the Orion nebula (Rubin et al. 1991). 
We obtain $\rm <O^{2+}/O>/<S^{2+}/S>\simeq 0.04$ and 0.09 towards the Off 
and On position respectively.
The variations of $\rm <O^{2+}/O>/<S^{2+}/S>$ versus the effective 
stellar temperature $\rm  T_{eff}$ for ionizing luminosities in the range 
$\rm log N_L= 49-50$, indicates $\rm T_{eff}\simeq 35500$ (Fig.~8 in
Simpson et al. 1995). An effective temperature 
of 38000~K, typical of an O7 V star, requires a ratio 
$\rm <O^{2+}/O>/<S^{2+}/S>$ of a few, i.e. 10 to 20 times larger. 
There is a priori a restriction in directly applying the models of Rubin
since the density of the Trifid (100\cmmt) is somewhat less than the
density considered in the calculations. 
However, in the 
range of temperatures and Ly-continuum luminosities considered, the ratio 
of the ionic lines is almost independent of the density in the range 
$100-1000\cmmt$ (Rubin, 1994). 

Therefore, the effective stellar temperature of the central star exciting
the Trifid is $\rm T_{eff}\simeq 35500\K$. Interestingly, this value 
is typical of an O7.5~III star, a spectral type proposed by Walborn (1973)
for HD ~164492A, and not an O7~V star (the classification proposed by Lynds
et al. 1985). In both cases, 
the physical properties of the exciting star are very similar,
and, in particular the number of Lyman-Continuum photons varies by 
$\approx 30\%$, reaching $9.6\times 10^{48}\smu$ for an 07.5~III
\footnote{There is an error in Table~II of the article of Panagia (1973)~:
the correct number of Ly-c photons for an 07.5~III star is not 49.98, as 
mentionned, but 48.98, as can be checked from the excitation 
parameter U in the following column.}.

\subsection{The FUV field Intensity}

A simple estimate of the FUV field intensity can be obtained for spherical, 
ionization-bounded, \Hp\
regions excited by standard O stars and in pressure equilibrium with the 
parent cloud (Spitzer, 1978). The FUV field intensity $\rm G_0$ 
is related to the hydrogen nuclei density $\rm n_H$ via 
$\rm G_0\simeq 3\times 10^5 \left( (n_H/10^4\cmmt)/C \right)^{1.33}$ 
(Spitzer, 1978). Here, $\rm C= n_H/n_e\sim 100 $ 
in an \Hp\ region in pressure equilibrium with the ambient cloud, 
$\rm n_e= 100\cmmt$ is the  mean electron density in the $\Hp$ 
region (Chaisson \& Willson, 1975) and 
$\rm G_0$ is measured in units of the average interstellar field 
flux of $1.6\times 10^{-3}\erg\smu\cmmd$ (Habing, 1968). 
One derives
a FUV field $\rm G_0\simeq 700-1700$ at the border of the \Hp\ region. 

A more accurate value is obtained by considering the infrared dust
continuum emission. If one makes the simple hypothesis that in the PDR
the dust is heated only by the incident FUV field, then, at first order, 
the dust temperature is related to $\rm G_0$ via 
$\rm T_d= 12.2 G_0^{0.2}~K$ (Hollenbach, Tielens \& Takahashi, 1991). 
The temperature estimated for the warm dust layer in TC2 ($\approx 46\K$;
see Sect.~6)
yields a FUV field intensity $\rm G_0 \simeq 800$. This value is in good 
agreement with the (crude) previous estimate. It is also consistent with the 
$\rm (\Oi~145/63\mu m)$ and  $\rm (\Oi~63/\Cp~158\mu m)$ line ratios
measured towards TC2, as we discuss in the following paragraphs. 
We will adopt $\rm G_0= 1000$ in the rest of the article.

\section{The Bright Rim and the Photoionized Envelope : zone II}

The Bright-Rimmed Globule TC2 is located on the Southern border of the \Hp\ 
region, at approximately $1\pc$ of the exciting star of the nebula~:
HD~164492A. An optical $\rm H\alpha$ image of the nebula (taken from CL98),
is shown Fig.~2. The TC2 region is marked by a white rectangle. The
globule exhibits the typical bright rim of that class of photoionized 
condensations (see also the magnified view in Fig.~4); it indicates 
that the condensation lies almost in the plane of the sky, 
coplanar with HD~164492A, on the Southern border of the nebula. 
Hence, the projected distance must be close to the physical distance
between both objects. We determined the mean radius $\rm R_g$ 
of TC2, based on the optical image~: $\rm R_g\simeq 4\times 10^{17}\cm$. 

Some weak extended [S{\small II}] emission is detected South 
of the bright rim (Fig.~2). The brightness is much higher at the surface 
of the
globule than in the dust lanes of the nebula. This indicates the presence
of ionized gas between the globule and the observer and gives direct
evidence of the physical association between TC2 and the Trifid. The
kinematical analysis of the molecular gas shows that the globule belongs 
to a large-scale feature connected to the dust lanes on the front 
side, and does not lie on the back side of the nebula (Lefloch et
al. 2002). Therefore TC2 is ``bathing'' in the nebular ionized gas. 

The peak of brightness is found, as expected, in the 
direction of the ionizing star of the nebula HD~164492A. However, the 
major axis of the globule does not point towards the star but in the North 
direction. This difference in the orientations might reflect some 
inhomogeneties in the gas distribution or the influence of other 
stars which formed at the same time as HD~164492A~: the energy released 
would have affected the expansion of the \Hp\ region. 
Indeed, Lefloch et al. (2001)  have reported the presence of several
young high- and intermediate-mass stars in the \Hp\ region,  
some of which have been identified as strong X-ray emitters 
(Rho et al. 2001). 

On the left side of the bright rim, close to the offset position
$(+75\arcsec, -95\arcsec)$, the HH~399 jet powered by TC2 (Figs.~4-5b) 
is detected 
as a conspicuous bright filament which seems to penetrate the surface 
and makes a P.A. of $\sim 20\deg$ with respect to North.
At the basis of the jet, a ``dark filament'' is detected in absorption 
against the bright rim and  emerges from the globule. This
filament is probably related to some inhomogeneities or some instabilities 
at the surface of the globule. 

Depending on the spectral 
type adopted (see Sect.~3.2) the Lyman-continuum (Ly-c) luminosity
of the star lies in the range
$7-10\times 10^{48}\smu$. Moreover, a fraction of the ionizing photons is
consumed in radiative recombinations within the ionized gas or absorbed by
the dust between the exciting star and the globule. 
The dusty aspect of the nebula in the optical images suggests
that the absorption of the ionizing radiation might play a non-negligible
role. CL98 presented a map of the free-free emission at 20~cm, obtained
with the VLA at $10\arcsec$ resolution. We show in Fig.~5a an excerpt of
this map, centered on the globule. The map reveals mainly the large-scale
emission from the nebula. The photoionized layer is also detected as a
region of maximum flux on the western side of the globule. However, the
angular resolution is not high enough to discriminate the emission of the
ionization front and the ionized gas close to the surface from the
nebular emission, associated with lower-density gas. 

In this section, we first evaluate the electron density and the intensity
of the ionizing field at the surface of the globule from the free-free
emission observed at the VLA. We then use the various fine-structure ionic
lines detected with ISO to determine the geometry of the illumination with
respect to the globule.

\subsection{The Bright Rim}

In order to characterize
the ionizing conditions at the surface of TC2 (ionizing 
field intensity, electron density),  we have 
studied the emission of the bright rim with the VLA at high angular 
resolution ($\sim 1.5\arcsec$). 
The map of the 3.6cm continuum emission is shown in Fig.~5b. 

The bright rim is clearly detected; it is 
marginally resolved in its transverse direction with a thickness 
$\approx 1.5-2.0\arcsec= (3.5-5)\times 10^{16}\cm$. 
Integrating over the bright rim, we derive a total flux of 
$2.4\pm 0.2 \mjy$.  The photo-ionization of the surface layers induces the 
formation of a shell of ionized gas around the globule, which is 
detected in the optical as the bright rim. The 
radiative recombinations which take place in the shell shield the molecular
material of the globule from the incoming ionizing flux.
Previous theoretical work (Bertoldi, 1989; Lefloch \& Lazareff 1994)
showed that the ionized shell can be modeled as a shell of uniform 
density $\rm n_e$
and thickness $\eta \rm R_g$~: $\rm n_e$ is the ionized gas density at the 
surface of the globule, $\rm R_g$ is the radius of the globule, essentially
the radius of the cross-section to the ionizing field, and $\eta$ is a 
geometrical factor of the order 0.1-0.2. For standard ionization conditions
most of the impinging ionizing photons are consumed in the shell. 
The free-free radiation flux 
across the ionized shell detected in an area of size $\theta_s$ 
can be expressed as a function of the electronic density and temperature, 
and $\theta_s$ (Lefloch et al. 1997)~:

\begin{eqnarray}
\left(\frac{S_{\nu}}{1\mjy}\right) & = & 2.36\times 10^{-5} 
\left(\frac{n_e}{1\cmmt}\right)^2 \left(\frac{\eta \rm R_g}{0.1\pc}\right)
\nonumber \\
 & & \times   \left(\frac{T_e}{10^4\K}\right)^{-0.35} 
\left(\frac{\nu}{1\ghz}\right)^{-0.1} 
\left(\frac{\theta_s}{10\arcsec}\right)^{2} \nonumber 
\end{eqnarray}
\smallskip

We assume in this formula that the size of the emitting region along 
the line of sight is of the same order as the thickness of the ionized
shell, i.e. we neglect any inclination effect. 
We adopted the mean electron temperature in the nebula determined by 
Chaisson \& Willson (1975) from centimeter continuum measurements at 
$4\arcmin$ resolution~: $\rm T_e\simeq 8150\K$. 
We measure a total flux of $2.3\mjy$  in an area 
of 74 arcsec$^2$, defined by the contour level at 
$20\mujy\, \rm beam^{-1}$. 
Assuming that the bright rim radial size 
is similiar to the thickness of the ionized shell, i.e., $\eta=  0.08-0.11$. 
We derive an average electron density $\rm n_e= (1.0-1.2)\times 10^3\cmmt$ 
over the bright rim. At the brightness peak,  the 3.6~cm flux is 
$2.6\times 10^{-4}\jy\, \rm beam^{-1}$, hence
the electron density reaches a local maximum of $n_e= 1800\cmmt$ over the
telescope beam. 

The effective flux of Ly-c photons ionizing fresh
material is $n_e c_i\simeq 10^9\cmmd\smu$ ($c_i\simeq 10\kms$ is
the ionized gas sound speed). As expected, this is much less 
than the flux of photons consumed in the photoionized layer, which is 
equal to $\alpha_B \eta n_e^2 R_g= 1.4\times 10^{10}\cmmd\smu$ 
($\alpha_B= 2.7\times 10^{-13}\cmpt\smu$
is the recombination coefficient on hydrogen levels $n \geq 2$). 
Taking into account the attenuation of the radiation between the star
and the globule by the ionized gas, this value is consistent
with the (absorption free) value derived from the Ly-c luminosity of the 
exciting star, $8\times 10^{10}\cmmd\smu$ at the projected distance of 
TC2. 

\subsection{Fine-Structure Atomic Line Emission}

We detected with the SWS spectrometer the following 
ionic and fine-structure atomic lines~: $\Nep$~$\rm 12.7\mu m$, 
$\Nepp$~$\rm 15.5 \mu m$, $\sulfpp$~$\rm 18.7 \mu m$, 
$\sulfpp$~$\rm 33.5 \mu m$ and $\Sip$~$\rm 34.8\mu m$. 
The spectra are displayed in Fig.~6. 
The fluxes of the lines identified are given in Table~\ref{flux_sws}.
A priori, several regions with very different physical conditions
contribute to the emission detected~: the $\Hp$ region (see Sect.~3), 
the bright rim and the PDR (in the case of $\Sip$). 
Since the $\Sip$~$\rm 34.8\mu m$ is believed to arise from the PDR, 
it is discussed below in Sect.~5.4. 
The other lines have been discussed in Sect.~3 and we concluded that
within the SWS beam and the observational uncertainties the emission
is dominated by the HII region.

In order to
precise the spatial distribution of the $\Nep$ and $\Nepp$ lines, we use 
the CVF data, obtained with a much higher angular resolution
($4\arcsec - 6\arcsec$). Moreover, the CVF data benefits a better SNR 
because of the 
low spectral resolution ($\approx 40$).
In order to determine the emission from the 
globule itself, we have subtracted the contribution of the \Hp\ region,
estimated from a nearby reference position (position E, Fig.~7). 
The emission from the globule, once the contribution of 
the nebular gas was removed, is displayed in Fig.~7. 
The $\Nep$ emission is mostly detected along the border of the globule,
at the surface, slightly shifted outside with respect to the emission of
the PDR. This spatial shift is shown in Fig.~7 where the contours of the
PAH $\rm 11.3\mu m$ band are superposed on the $\Nep$ emission map.
The emission delineates a thin layer of $\simeq 5\arcsec$. 
It can be seen 
that {\em there is no $\Nep$ emission detected over the body of 
the globule}. A spectrum at position F in  the main body of TC2 
(Fig.~8f) and confirms the absence of emission longwards of 
$\rm 8\mum$. Therefore, {\em the front side of the globule is hardly 
illuminated by the Ly-c photons from the exciting star. In particular, 
the main body of the globule is not photoionized}. 
Some weak $\Nep$ emission is detected all around the globule, 
especially in
the Northerwestern direction, i.e. towards the ionizing stars. 
We suggest that it could trace the freely expanding layers of
the photoionized envelope around TC2. This gas is expected to leave the
surface of TC2 with a velocity close to the ionized gas sound speed, $\sim
10\kms$ (LL94). The detection of high gas velocities, from spectroscopic 
measurements, and the study of their spatial
distribution  would allow to confirm the association of this  
component with the photoevaporated envelope of the globule. 

The distribution of the $\Nepp$ $\rm 15.5\mu m$ emission was obtained 
following the same procedure, and is displayed in Fig.~7. The $\Nepp$ 
map looks much noisier than $\Nep$. 
Again, there is no emission detected towards the body of the 
globule. The  $\Nepp$ emission is limited to the Northwestern quadrant, 
between the globule and the ionizing stars. It overlaps 
with the outer part of the $\Nep$ flux distribution, at the surface of the 
globule, and in the extended component which possibly traces the
photoevaporated envelope. This is consistent with the high ionization
potential of the $\Nepp$ line (41~eV) with respect to to the $\Nep$ line 
(22~eV). 

Additional lines were detected in the far-infrared with the LWS. 
The emission from TC2 was estimated by subtracting the 
emission of the reference position (see Sect.~3) to the spectrum 
towards the globule. In this case, no correcting factor 
was applied to the fluxes since TC2 fills only half the LWS beam
(Fig.~4). We observe a marked increase of  
the density around the globule. The ratio ($\Opp ~52/88)$ ratio 
yields $\rm n_e= 170\cmmt$. A more pronounced 
increase is observed using the ($\Npp ~57/\Opp ~52$) ratio~: 
$\rm n_e= 550\cmmt$, when adopting an abundance ratio of 0.5. 

One cannot exclude that the contribution of the \Hp\ region along the line of 
sight of TC2 is somewhat underestimated by the  procedure descrived in
Sect.~3. 
On the other hand, an additional factor can account for the observed
increase in density~: the photoionized envelope 
of the globule. As discussed in Sect.~4.1, the latter can be modeled as a 
layer of effective thickness
$5\times 10^{16}\cm$ and density $\rm n_e= (1-2)\times 10^3\cmmt$, 
which expands almost freely around the globule. The emission of this layer 
could account for the observed increase in density towards TC2 and for the 
discrepancy between the density estimates from the ($\Opp ~52/88)$ and
($\Npp ~57/\Opp ~52$) ratios. 
This is because the $\Opp$ $\rm 52\mu m$
and $\Npp$ $\rm 57\mu m$ lines are more sensitive to dense material
than the $\Opp$ $\rm 88\mu m$ line (their critical densities are 
similar to the ionization front density 
whereas it is only $\sim 460\cmmt$ for $\Opp$ $\rm 88\mu m$). 
Indeed, a simple modelling with two layers of density $200\cmmt$ and 
$2000\cmmt$, thickness $\rm \ell = 1.3\times 10^{18}\cm$ and 
$5\times 10^{16}\cmmt$ respectively can account for the emission
observed. The contribution of the dense layer is almost unnoticed at 
$\rm 88\mu m$ whereas it constitutes $20-30\%$ of the observed flux for 
the $\rm 57 \mu m$ and $\rm 52 \mu m$ lines respectively. 

\subsection{Summary of the SWS and LWS Observations}

The analysis of the various fine structure atomic lines  detected with the 
SWS and the LWS spectrometers confirms that the bright-rimmed globule is 
bathing in the nebular gas of density $\rm n_e= 50-100\cmmt$. For most
of the fine-structure atomic lines, the nebular gas dominates the
contribution of the photoionized layer surrounding the globule.
The photoionized gas envelope is detected at the VLA; it is 
indirectly detected in  the $\Npp$ $\rm 57\mu m$ and $\Opp$ 
$\rm 52 \mu m$ lines. The surface density is $\simeq 1000\cmmt$ and rises
up to $\rm 2000\cmmt$. 
The photoionized envelope of TC2 is also detected in the $\Nepp$ and 
$\Nep$ lines. The mapping of these lines shows that 
the globule is illuminated on the rear side. 

\section{The Photon-Dominated Region : zone III}

In this section we characterize the properties of the PDR~: geometry, 
hydrogen density and gas column density. 
We first study the geometry of the PDR 
from the emission of the Unidentified Infrared Bands (UIBs) at 
6.2, 7.7, 8.6 and $\rm 11.3\mu m$ as observed with ISOCAM. The UIBs are 
a good tracer of the strong UV field region associated with a PDR. 
Although the exact chemical composition of these bands 
is still not known, the best candidates appear to be the 
Polycyclic Aromatic Hydrocarbons, or PAHs (Puget \& L\'eger 1989).
We will use 
indistinctly the former or the latter denomination in what follows. 
The physical conditions in the PDR are quantified
from the emission of dust and of the mid-infrared lines detected 
with LWS and SWS~: $\Sip$~$\rm 34.8 \mu m$, $\Oi$~$\rm 63 \mu m$, 
$\Oi$~$\rm 145 \mu m$ and $\rm \Cp$~$\rm 158 \mu m$.  
The $\Oi$ and $\Cp$ lines 
are especially interesting tools in the study of PDRs since they can be 
used to probe the excitation conditions in the gas. 
The line fluxes are compared 
with the recent models of Kaufman et al. (1999; hereafter K99) and 
 Wolfire, Tielens \& Hollenbach (1990).
These models extend the work by Tielens \& Hollenbach (1985), 
Hollenbach, Takahashi \& Tielens (1991)
The main difference is that the
model of K99 use very recent grain photoelectric heating rates, which
include treaments of small grains and large molecules (PAHs).
We estimated that such a contribution might be
important in the case of TC2 since our CVF data show that the infrared
emission shortwards of $15\mu m$ is dominated by the PAHs emission
bands.

\subsection{The Emission of the UIBs}

As can be seen in Fig.~8, the UIBs (or PAHs) bands at 
6.2, 7.7, 8.6 and 11.3$\rm \mu m$ dominates the 
the spectral emission in the mid-infrared range $\rm 5-17\mu m$, 
as observed with the CVF.  
One notes also in the spectra a strong 12.7$\mu$m line that is 
a combination of the [NeII] line and of a PAH band, and the 
presence of a continuum emission at wavelength greater than 13$\mu$m, 
also well detected in the SWS spectra 
(see Sect.~6.2), and the presence of the [NeIII] 15.5 $\mu$m line.
To study the spatial distribution of the various features in the CVF spectra, 
we have made a 
spectral decomposition using a Gaussian function for the [NeII] line, 
Lorentzian lines for the PAHs bands 
(Boulanger et al. 1998) and a grey body for the continuum emission.
The resulting maps are shown in Fig.~7.  

All the UIBs are unambiguously 
detected and have the same spatial distribution, following closely the
border of the globule. We note that there is no spatial 
segregation between the PAHs bands probably as a consequence of the 
remote distance to the Trifid nebula.
The emission is shifted by $\approx 2\arcsec$ towards the interior of the 
globule with respect to the photoionized region, as traced by the 
$\rm \Nep$ line (Fig.~7). 

The emission shows a local minimum in the center of the globule. 
The flux distribution is the brightest in the PDR and reaches its maximum
close to the offset position $(+65\arcsec, -125\arcsec)$. 
The PAH emitting region in the PDR is mostly unresolved at all 
wavelengths; this prevents any detailed physical analysis of the 
latter. 
The dark filament absorbs a fraction of the infrared radiation coming from 
the PDR, causing the presence of a local minimum in the brightness 
distribution. Because of the weakness of the radiation field, almost no
emission at all is detected over the body of the globule. 
Only a weak feature is detected around $\rm 7.7 \mu m$
(Fig.~8). Some weak emission is also detected outside the globule, 
behind the ionization front (Fig.~7). 

We now compare the relative variations of the PAH bands
measured at a few positions over the globule and the  $\Hp$ region.  
In order to increase the SNR, the emission has been averaged over four
pixels at each position. The resulting spectra are displayed in 
Fig.~8. The contribution of the nebula, as estimated from the reference 
position (Fig.~8e) has been subtracted to the spectra in order to outline the 
emission of the globule. The almost flat spectrum obtained in the core TC2,
especially the absence of ionic lines either in absorption or in emission, 
shows that the nebular gas emission was removed properly. 
It is also consistent with TC2 being heated on the rear side. 

We concentrate on the 7.7 and 11.3$\mu$m as they have the highest SNR.  
The bands have a ratio 7.7/11.3$\simeq 0.9$ in the PDR. 
This ratio is $\sim$1.2 at the reference position (panel e) in the 
\Hp\ region and it reaches $\approx 2$ in the direction of the ionizing star.
Part of this variation
of the 7.7/11.3 ratio is probably related to variations in the charge of
the PAHs.
The current models and laboratory experiments on the PAHs emission 
show that the relative intensity of the various bands are strongly 
sensitive to the ionization state. In particular the intensities of the 
C-C stretching modes (6.2 and $\rm 7.7\mu m$) and the C-H in-plane 
bending mode ($\rm 8.6\mu m$) are generally stronger in PAH cations than in 
PAH neutrals by a factor of 10 (see e.g. Joblin et al. 1996), whereas the 
intensity of the C-H out-of-plane bending mode ($\rm 11.3\mu m$) seems to 
decrease with the ionization. 

In the PDR of the globule, the 
models of Bakes \& Tielens (1998) and  Dartois \& d'Hendecourt (1997)
predict a very small fraction of PAH cations (0.01 or less). 
In the  photoionized gas around TC2 on the contrary, 
the spectra look very similar to those of a mix of PAH cations 
superposed to a weak continuum (see Fig.~3d in Allamandola, Hudgins 
\& Sandford, 1999). 
Therefore the decrease of the $\rm 11.3\mu m$
band can be understood as the dehydrogenation and photoionization 
of the smallest PAHs exposed to the strong UV field of the \Hp\ region. 
A similar effect has been reported 
in the \Hp\ region M17 by Verstraete et al. (1996) and Crete et al. (1999) 
and in NGC~1333 by Joblin et al. (1996). 

However, we observe a marked increase of the 7.7/11.3 ratio, above 2, inside
the globule. This value is comparable to that observed in the ionized gas,
whereas the ionizing field at the surface of the globule, as traced by 
the $\Nep$ and the $\Nepp$ lines, 
is much lower. Hence, the relative variation of the 7.7/11.3 ratio cannot be
directly accounted for by standard models of PAH excitation.
As we discuss below, {\em the PAH emission observed towards the main body of
the globule actually comes from the PDR located on the rear side} and has
suffered strong absorption from the globule material (the millimeter dust
map indicates an average $\rm A_v= 20$ over the body of the globule).
The absorption by the silicates is so large
that hardly any radiation escapes from the globule longwards of $8\mu m$.
This effect is illustrated by the CVF spectrum taken
at position F, in the main body of TC2. Almost no emission at all is
detected in the spectral window, apart from a weak feature coinciding
with the PAH band at $7.7\mu m$.

It is interesting  to compare the emission maps of the globule at $7.7\mu m$
and in the $\Nep$ line at $12.7\mu m$. In both maps, there is no emission in
the Northeastern region. This is the region where the reference spectrum
was taken (position E, Fig.~7). The border of the globule looks
very bright at both wavelengths. Whereas the main body of TC2 appears
void of $\Nep$ emission, as mentionned before, it is still radiating some
flux at $7.7\mu m$ (see Fig.~7). Cernicharo et al. (2000) showed that at
5.3, 6.6 and $7.5\mu m$ there is a narrow window in which the absorption by
 the ices and the silicates is low enough 
to let the radiation escape even from the deeply embedded cores of 
Class 0 protostars. 

In order to test our hypothesis, we have calculated the spectrum of the 
radiation emerging at the front side of the globule, assuming the 
emitting region (the PDR) to be located 
at the rear surface of the globule. The extinction through the 
globule was computed from an empirical absorption profile constructed by 
Cernicharo et al. (2000) and scaled with the visual extinction inside TC2. 
This empirical profile is based on the CVF 
spectrum of the Class I protostar VLA4 in the L1641 molecular cloud in 
Orion. It takes into account the contribution of silicates,  
the ices of methanol $\rm CH_3OH$, water $\rm H_2O$ and carbon dioxide
$\rm CO_2$. 
As noted by Cernicharo et al. (2000), the spectrum of VLA4 is very similar
to the spectra observed towards deeply embedded objects in massive 
star-forming regions. We also note that VLA4 and TC2 are in a similar 
evolutionary stage. Hence, we believe that despite some possible abundance 
variations 
between both protostellar cores, the VLA4 absorption profile should 
represent a good approximation to the absorption profile in TC2. 

We have applied this absorption profile to the spectrum of the bright PDR
(position D) for a visual extinction $\rm A_v= 20$, a value derived from the
1.3mm continuum map, and similar to the average extinction over the globule.
We show the calculated spectrum in Fig.~9.  The $7.7\mu m$
band intensity is $\sim 70$ MJy/sr whereas the intensity of the 
$\rm 11.3\mu m$ band and the $\Nep$ line is now only $\sim 40$ MJy/sr. 
The synthetic spectrum of the PDR appears very similar 
to the spectrum in the body of the globule at position F. 
The apparent strong enhancement of the $7.7\mu m$ band with respect to 
the $11.3\mu m$ band is mainly a selective absorption effect of 
material located on the rear side of the globule. 

To summarize, the PAH bands at 6.2, 7.7, 8.6 and $\rm 11.3\mu m$ are
unambigusouly detected in TC2. The variations of their relative intensities
are in agreement with the models of PAH excitation. 
They show evidence for the main body of TC2 being illuminated on the rear 
side and not on the front side. This is in agreement with the analysis 
of the $\Nep$ and $\Nepp$ lines. As a consequence, the line intensities
can be obtained, at first order, from the integrated fluxes by 
dividing by the area of the globule encompassed by the telescope beam.

\subsection{$\Cp$ line emission in the globule}

The LWS spectra of the $\Cp$ and $\Oi$ lines are shown in Fig.~6. 
The emission from the globule was obtained by subtracting a reference
position to the spectra taken towards TC2 (see Sect.~2.5). 
The uncertainties in the flux of the $\Oi$ lines are rather weak 
since the emission at the reference position appears much weaker than
towards the globule. On the
contrary, the variations of the $\Cp$ line are much weaker and the
contribution of TC2 amounts to $\approx 25\%$ of the total flux 
(see Table~\ref{flux_lws}). The procedure applied to estimate the 
emission from the globule ignores a possible contribution of the 
photoionized envelope surrounding TC2 to the observed flux. 
Indeed, the analysis of the far-infrared $\sulfpp$ and $\Opp$ lines 
carried out in Sect.~3.1 suggested
that this region was contributing to the flux detected. 
The $\Cp$ line is one of the main tools used to probe the physical conditions
in PDRs as it provides a direct estimate of the mass and column density
of PDR gas. 
It is therefore important to estimate as accurately as possible 
the $\Cp$ flux coming from the PDR itself, and leave aside any other
contribution to the flux measured. 

We consider the 20~cm free-free emission map of the Trifid presented 
in CL98 (Fig.~5a). 
These data were obtained at a resolution of $10\arcsec$. This allows
to trace the extended emission from the main body of the globule,
unlike the high-angular observations presented in Sect.~3.1.
We apply the method described by Heiles (1994), who showed that 
intensity $\rm I_{C}$ of the $\Cp$ line scales with the 
brightness temperature of the free-free emission in low-density 
region  and we  derive the relation~: 
\begin{equation}
\frac{I_C}{T_{B,20cm}} = 0.8 \delta_{C} \K^{-1}
\end{equation}
where $\rm I_{C}$ is expressed in units of
$\erg\cmmd\smu\srmu$ and $\delta_C$ is the ionic abundance
$\rm C^{+}/H^{+}$ in units of $3\times 10^{-4}$. The exact ionic
abundance of $\rm C^{+}$ depends on various factors~: density, effective
stellar temperature. Based on the models of Rubin (1985), we estimate 
$\delta_C$ to range between 0.5 and 0.8. 

The free-fre emission map shows some extended emission in addition to the
bright rim. The emission peaks at $\rm 33\mjy/beam $ in the bright rim.
From the distribution of the contour levels around the globule, we find
that the average intensity in the surrounding nebular gas is
$\rm 20\mjy/beam$. This implies a peak flux of $\rm 13\mjy/beam$ in the
bright rim. Following the same method as in Sect.~3.1, we infer a local
density $\rm n_e= 1900\cmmt$, in good agreement with our measurement at
higher angular resolution. After subtracting the contribution of the
nebular gas and integrating the emission over the globule, we obtain a
mean brightness temperature $\rm T_{B,20cm}= 4\K$ for the residual
emission. Since the contribution of the nebular gas was removed,
this residual is the emission of the photoionized envelope and the
ionization front, i.e. the bright rim. Hence, the intensity of the
$\Cp$ line $\rm I_C\simeq 1.5\times 10^{-4}\erg\cmmd\smu\srmu$.
This value represents approximately $30\%$ of the flux detected with the 
LWS towards the globule. After correcting for the contribution of the 
surface ionized layers, we obtain that the flux of the $\Cp$ line 
in the PDR gas is 
$\simeq 1.8\times 10^{-11}\erg\cmmd\smu$. 


In the optically thin limit, the mass of atomic gas can be 
easily obtained from the observed $\Cp$ line flux, using the analytic 
formula derived by Wolfire et al. (1990)~: 
\begin{eqnarray}
M_a &= & 5.8 \left(\frac{d}{1\kpc}\right)^2 
\left[\frac{1.4\times 10^{-4}}{x(\rm CII)}\right] 
\left(\frac{F_{\rm CII}}{10^{-17}W\cmmd}\right)  \nonumber \\
    & & \times 
\left(\frac{10^{-21}\erg\cmmd\smu\srmu \rm atom^{-1}}{\Lambda(\rm CII)}\right)
   \msol\nonumber
\end{eqnarray}
where $\Lambda(\rm CII)$ is the cooling rate in the $\Cp$~$158\mu m$ line
per carbon atom and per steradian, $x(\rm CII)$ is the abundance of ionized 
carbon per hydrogen atom. 
Here, we adopt a carbon abundance $x(\rm CII)= 1.4\times 10^{-4}$ (see
K99). 
The atomic gas temperature in the PDR of TC2 is close to 300~K (see next 
section).  
For hydrogen densities of $10^4\cmmt$, comparable to that in the PDR gas,
the cooling rate $\Lambda(\rm CII)$ is then 
$0.8-1.0\times 10^{-21}\erg\cmmd\srmu \rm atom^{-1}$. The mass of atomic gas
in the PDR is therefore $\rm M_a= 2.9-3.7\msol$. From the fraction of the 
globule's area that fills the LWS beam, we derive the average hydrogen 
column density of PDR gas~: $\rm N(H)= (1.5-1.9)\times 10^{21}\cmmd$. 

\subsection{Physical Conditions in the PDR from the {\rm $\Oi$} and 
  {\rm $\Cp$} lines}

We present below the physical conditions derived from the analysis of the 
far-infrared lines and based on the model of K99. We find a good 
agreement between this model and the observational data. 
The physical parameters of the globule are summarized in Table~5.  

Based on the FUV field intensity $\rm G_0\simeq 1000$ and using the model
of K99, we find that the gas is heated by 
photo-electrons to a temperature  of $\simeq 300~K$ at the surface of the 
PDR,  a value which actually depends rather little on the  actual density 
in the PDR (see their Fig.~2). This value is also a good estimate of the 
actual temperature in the PDR.

The hydrogen density in the PDR was estimated 
from a Large-Velocity Gradient analysis of the  $\Oi$ lines. 
This approach offers the advantage of being independent of 
the oxygen elemental abundance. The size of the emitting region in the LWS
beam is taken to be $45\arcsec$. We took $1.3\kms$ as linewidth, based on
the observations of the millimeter lines of $\hcop$ and CS in the head of 
the globule (Sect.~7.2). The core density derived from the
millimeter dust thermal emission provides an upper limit to the density in
the PDR~: $n(\htwo)= 3\times 10^5\cmmt$. Hence, in order to account for the
observed $\Oi$ fluxes at $\rm 63\mu m$ and $\rm 145\mu m$, the gas
temperature has to be larger than $200\K$. The best match is obtained for 
a column density $\rm N(O)= 7 \times 10^{17}\cmmd$ (Fig.~10). At a 
temperature of 
$300\K$, as derived above for the PDR, the molecular hydrogen density is 
$\rm n(\htwo)= 6\times 10^4\cmmt$. Adopting an oxygen elemental abundance
$\rm [O]/[H]= 3.0\times 10^{-4}$, the corresponding gas column density in
the PDR is $\approx 2\times 10^{21}\cmmd$.

There is a possible bias in this analysis as the globule is 
illuminated under some inclination angle while it is viewed face-on by us. 
This anisotropy in the illumination drives heterogeneous conditions across the
globule (i.e. in the plane of the sky). As a consequence, the filling
factor of the $\Oi$~$\rm 63 \mu m$ line, more easily excited, is
likely to be larger than that of the $\rm 145\mu m$ line. 
This effect was not taken into account in the present calculation and
results in an underestimate of the $\Oi$ column density whereas the
hydrogen density is overestimated. 

The $\rm (\Oi~63/\Cp~158)$ 
and $\rm (\Oi 145/63)$ line ratios provide another method to derive 
the parameters of the PDR. These ratios are 0.066 and 4.2 respectively; 
using Figs.~4-5 in K99, we obtain 
direct estimates of the FUV field intensity $\rm G_0= 1000$ and 
of the hydrogen nuclei density $n(H)= 2\times 10^4\cmmt$. This estimate of
$\rm G_0$ is in good agreement with the previous, independent,
determination, based on the far-infrared dust continuum. 
In the range of values taken by the $\rm (\Oi~63/\Cp~158)$ 
ratio, the ratio is not very sensitive to the FUV field intensity 
$\rm G_0$. 
Hence, an overestimate of the $\Cp$ line flux would imply that the actual 
PDR density is somewhat larger. An upper limit of 10 on this ratio yields 
a hydrogen {\em nuclei} density of $\sim 3\times 10^4\cmmt$. 
We determine the electron density in the PDR from the photoelectric heating
efficiency $\epsilon$, which is governed by the factor 
$\rm G_0 T^{1/2} n_e^{-1}$ (see e.g. Bakes \& Tielens 1994). The integrated
intensity of all the FIR lines is well approximated by the sum of the
$\Oi$~$\rm 63\mu m$ and $\rm \Cp ~158\mu m$ intensities, which amounts to 
$2.6\times 10^{-3}\erg\smu\cmmd\srmu$. We note that since the cooling of 
the PDR gas is dominated by the $\Oi$~$\rm 63\mu m$ line, the uncertainties
in the $\Cp$ line flux do not affect significantly the total cooling nor 
the conditions of thermal balance. We estimate the integrated FIR continuum
intensity radiated by the PDR from the warm component determined in the 
spectral energy distribution~: $0.21\erg\smu\cmmd\srmu$. On the other hand,
the FUV photon heating amounts to 
$\approx 1.6\times 10^{-3}G_0/(4\pi)= 0.102 \erg\smu\cmmd\srmu$. This 
value represents a lower limit to the total heating since the contribution 
of the photons outside the FUV band has been neglected here and could 
represent a substantial fraction (up to 1) of the FUV heating (Tielens 
and Hollenbach, 1985; Wolfire et al. 1989). The agreement between the FIR 
continuum radiation and the UV photon heating is therefore satisfactory and
supports our view that the warm dust layer detected with ISO/LWS and SWS 
(see Sect.~6) is indeed tracing the PDR at the surface of the globule. This
yields a photoelectric efficiency $\epsilon= 0.013$. From the variations 
of $\epsilon$ with the factor $\rm G_0 T^{1/2} n_e^{-1}$ (see Fig.~21 
of Bakes \& Tielens 1994), we estimate 
$\rm G_0 T^{1/2} n_e^{-1}\simeq 5\times 10^3 K^{1/2}\cmmt$, hence the 
electron density in the PDR~: $\rm n_e= 3.5\cmmt$. In the PDR, the
ionization fraction is determined by the photoionization of carbon. 
Adopting a carbon abundance of $1.4\times 10^{-4}$, we obtain an estimate 
of the  hydrogen density $\rm n(\htwo)= 1.3\times 10^{4}\cmmt$. 

\subsection{The Emission of the $\Sip$~ $\rm 34.8\mu m$ line }

The emission of the $\Sip$~$\rm 34.8\mu m$ line can be produced in the 
PDR of molecular clouds which are illuminated by strong FUV fields
(Tielens \& Hollenbach 1985) and also in $\Hp$ regions (Rubin, 1985). 
In the case of Orion, as prototype of massive star forming region, 
Walmsley et al. (1999) showed that the $\Sip$ emission
comes mainly from the PDR and not from the $\Hp$ region or the ionization 
front region.

We try to determine the origin of the emission observed towards the globule. 
It is difficult to draw any definite conclusion because of the 
lack of angular resolution, hence of information about the spatial 
distribution of the emission. Also, the noise in this band of the SWS
is known to be high, so that the actual variations between the three 
positions might be somewhat more  pronounced.
An important constraint comes from the  measurements of the thermal dust 
emission at 1.3mm; they indicate 
an average visual extinction $\rm A_v= 20$ across the globule (see Sect.~6). 
The values of the interstellar extinction tabulated by Mathis (1990)
allow to estimate the opacity of the globule at $\rm 35\mu m$~: 
$\tau_{35}\simeq 0.08$. Therefore {\em the globule is mainly transparent 
at this wavelength}. 

For the physical conditions encountered in the PDR of TC2,
the model of Wolfire, Tielens \& Hollenbach (1990)
predicts a typical value of 0.015 for the ($\Sip$~34.8/$\Oi$~63) ratio
whereas the observed ratio is 0.5.
Assuming all the emission arises from the PDR, this implies a
silicon gas phase abundance about 30 times higher than that assumed in their
computation, corresponding to $\rm [Si]/[H]\simeq 2.3\times10^{-5}$, 
i.e. 0.6 times the solar elemental silicon abundance
($\sim 3.6\times 10^{-5}$). This is unrealistically high
as observations in other $\Hp$ regions like Orion indicate that
about $90\%$ of Silicon is tied up in dust grains (Walmsely et al. 1999),
a value also similar to that found in diffuse interstellar clouds,
Therefore, the PDR of TC2 alone can not account for the measured 
$\Sip$ flux. 

Indeed, the SWS spectra show that the $\Sip$ flux at the Northern position
is only $\sim 15\%$ less than  at the Southern and Central positions,
whereas the globule fills only one third of the
SWS beam. Since the globule is transparent to the $\Sip$ radiation and it is
illuminated mostly on its rear side, one would rather expect a flux $\approx$
one third of that detected in the Central beam, i.e.
$\sim 3\times 10^{-19}\rm W \cmmd$. This is much less than the flux
detected. This reinforces our previous conclusion about the origin of the
$\Sip$ line.

Assuming the flux variation between the Central and Northern
position is due to the variation of PDR filling factor in the SWS beam,
one can
derive an estimate of the relative contribution of the PDR and the $\Hp$
region. The emitted flux is then $2.2\times 10^{-19}$ and
$7.4\times 10^{-19}\rm W\cmmd$  for the PDR and the $\Hp$ respectively; 
the emission of the PDR is somewhat dominated by the contribution of the $\Hp$ 
in the SWS beam. 
The model of Wolfire, Tielens \& Hollenbach (1990) is then consistent with
a gas phase
abundance of $6\times 10^{-6}$, i.e. $17\%$ the solar abundance.
This value is much more compatible with the determinations obtained
in other $\Hp$ regions. Nevetherless, we stress again that new observations 
with a better SNR are required to determine more precisely 
the silicon abundance in the PDR and the $\Hp$ region.  

\section{The Dust Continuum Emission}

\subsection{Warm Dust in the PDR~: zone III}

In this paragraph, we analyse the continuum emission detected with the SWS
in the range $\rm 2-45\mu m$. The full spectra obtained at the three
positions are shown in Fig.~11. 
The continuum emission becomes 
significant longwards of $\rm 30\mum$ in all the positions. 
We have fit the continuum using a black-body modified by a power law
$\tau_{\nu}\propto \nu^{\beta}$. A satisfactory fit was obtained 
for a dust temperature of 46~K, and a dust spectral index 
$\beta= 1.3$, typical of the values observed in the mid-infrared 
(Hildebrand, 1983). Here, we assume that the mid-IR flux originates from 
the PDR and fills the SWS beam. Following the reddening law 
determined by Lynds et al. (1985) for the Trifid, we estimate a hydrogen 
column density $\rm N(H)= 1.7\times 10^{21}\cmmd$. Adopting the standard
reddening law yields $\rm  N(H)= 2.9\times 10^{21}\cmmd$.
These values are only indicative since they rely on the geometry assumed 
for the emitting region. Reasonable fits can be obtained with 
the temperature in the range 43-48K, and hydrogen column densities of 
$(1-4)\times 10^{21}\cmmd$. 

The same procedure has been applied to the other positions observed
with the SWS. At the Northern position, the PDR of the globule
fills only partially the SWS beam. 
We have  estimated a size of $\approx 15\arcsec$ for the region 
encompassed by the beam. The other parameters for the warm 
layers were left identical otherwise. 
At the Southern position, the continuum emission could be fit with similar
parameters and assuming a smaller column density of warm dust 
($\rm N(H) \simeq 0.9\times10^{21}\cmmd$) at about the same temperature.  
The fits are shown superposed on the spectra in Fig.~11. 
They succeed rather well in reproducing the continuum flux longwards of 
$\rm 30\mu m$. 

It is therefore possible to explain the observed continuum 
emission  by a 
warm dust layer at about $46\K$ and a column density 
$\simeq 2-3\times 10^{21}\cmmd$. As we show below, such parameters also 
allow to account for
the continuum emission detected at longer wavelengths. 

\subsection{Hot Dust around TC2 : zones II and III}

Comparison between the fit of the warm dust component and the SWS data
shows actually the presence of a residual flux as a flat continuum 
between 10 and $\rm 30\mu m$. 
The continuum emission shortwards of $\rm 30\mu m$ is almost identical 
at the Northern and central positions. It is somewhat weaker  
in the South. 

The continuum emission is maximum  between 15 and $\rm 30\mu m$. 
Our CVF map of the continuum emission in the range $\rm 13 - 15\mu m$ 
 brings 
some more information on the distribution and the nature of this hot dust
component.  
The continuum emission at 15$\mu$m revealed from our spectral decomposition
is shown in Fig.~7. The map shows a good spatial correlation
with the $\rm 7.7\mu m$ band and the photoionized envelope. The continuum 
is brightest in the PDR. It is shifted 
outwards by $\approx 1-2\arcsec$ and gets stronger as one moves outside of 
the globule, in the photoevaporated gas. There is almost no emission from
the body of the globule. 
 We detect a higher continuum level in the Western side 
of the globule with respect to the Northern one. 
This difference might result from a higher
heating efficiency on the Western side because the UV photons arrive almost
normal to the surface. The rise of the continuum could also be related
to an increase of the gas density like what was observed in other PDRs  
(Abergel et al. 2002)
and in cirrus clouds (Miville-Desch\^enes et al. 2002). 
Within the present understanding of the nature of the mid-infrared emitters,
it is impossible to determine the origin of the increase of the continuum 
emission in denser regions but one plausible explanation could be related 
to a change of the dust size distribution.
There is also a rather good spatial correlation between the 15 $\mu$m continuum
and the distribution of the ionizing gas, as traced by the 
$\Nep$ line. 
The correlation of this emission with the photoionized envelope 
suggests that the emission detected probably comes from {\em very 
small grains} heated by the strong UV field.

\subsection{Cold Dust in TC2 : zone IV}

The thermal emission of the cold dust  was observed with 
the MPIfR 19 channel bolometer array at the IRAM 30m telescope (Fig.~3). 
We detected some weak emission, 
at typical fluxes of $5-15\mjy/ 15\arcsec \rm beam$, which spatially coincides 
with the region of high obscuration in the optical image, south of the 
ionization front. 
The millimeter continuum emission is sharply limited by the ionization front, 
and the  brightness contours are closely spaced between the bright rim and
the emission peak.   
The globule peaks at $160\mjy/11\arcsec\, \rm beam$ at the offset position
($70\arcsec$,$-124\arcsec$), only $15\arcsec$ behind the ionization front. 
From the contour at half power, we estimate a size
(beam-deconvolved) of $\approx 12\arcsec \times 32\arcsec$ 
($3.0\times 10^{17}\cm$ by $8.0\times10^{17}\cm$) for the core. 
Assuming an average dust temperature $\rm T_d= 20\K$, a  
spectral index $\beta= 2$ and an absorption coefficient
$\kappa_{250}= 0.1\cmpd\gmu$, we derive a hydrogen column 
density $\rm N(H)= 1.6\times 10^{23}\cmmd$ at the flux peak. Integrating 
over the contour at $80\mjy/11\arcsec \rm beam$  (HPFW), we find a total 
flux of $0.29\jy$ and estimate a core mass $\rm M= 27\msol$. This implies
a typical density $n(\htwo)= 3\times 10^{5}\cmmt$ for the core.

\subsection{The Spectral Energy Distribution}

The spectral energy distribution of the globule between 45 and 
$\rm 197\mu m$ was obtained by subtracting the  emission of the 
reference position to the LWS spectrum  taken towards TC2.  The size of the
emitting region
encompassed by the LWS beam is $\approx 45\arcsec$, based on the 
ISOCAM $12\mu m$ image and the cold dust millimeter emission (Fig.~3). 
We measured the 1.3mm continuum filling the LWS beam solid angle 
in order to constrain the fit to the cold dust component. It was found 
to be $\simeq 0.83$~Jy. The spectral energy distribution is shown in
Fig.~12. 
The emission could be satisfactorily fit by a two-component model~: 
a cold core of column density $\rm N(H)= 4.6\times 10^{22}\cmmd$ at a 
temperature $\rm T_d= 22\K$, with a dust spectral index $\beta= 2.0$, 
surrounded by a warm layer at a temperature $\rm T_d= 46\K$, 
with a hydrogen column density $\rm N(H)= 1.9\times 10^{21}\cmmd$, 
and a dust spectral index $\beta= 1.3$. 

The gas column density of the warm 
layer is not very high as it corresponds to $\rm A_v= 1-2$. We note 
that it is very similar to the column density of the PDR gas traced by
the $\Cp$~$\rm 158\mu m$ and the $\Oi$ lines. The warm layer detected with 
the LWS
also accounts for the mid-infrared continuum emission detected with the SWS 
longwards of $30\mu m$. Integrating under the fit of the warm dust 
component, and correcting for dilution in the LWS beam, we obtain
the infrared intensity radiated by the warm layer~: 
$\rm I_{IR}= 0.21\erg\smu\cmmd\srmu$.   
The temperature and the infrared luminosity of the warm layer are 
those expected for a PDR exposed to a FUV field $\rm G_0= 1000$. This is 
consistent with the intensity of the far-infrared $\Oi$ and $\Cp$ lines as 
observed with LWS (see above Sect.~5.3). The whole set of observational
data leads us to the conclusion that {\em the warm dust component detected 
with the LWS is actually tracing the PDR of the globule}. 
The average hydrogen column density is $\rm N(H)= 2.0\times 10^{21}\cmmd$. 
With a typical hydrogen nuclei density $n(H)= 2\times 10^4\cmmt$, as
estimated from the far-infrared line ratios (Sect.~5.3), we estimate 
the thickness of the PDR $\simeq 1.0\times 10^{17}\cm$. 

The cold dust component revealed in the spectral energy distribution 
is therefore tracing the innermost part of the globule (the core) which is
protected from the external UV radiation field. Hence, it seems a good
reasonable approximation to adopt 22~K as temperature of the dust core 
traced by the 1.3mm continuum emission in order to the estimate the 
mass of the core. This temperature is similar to those observed in the 
molecular cloud surrounding the nebula (Lefloch \& Cernicharo, 2000).
A more accurate determination of the mass and
column density of the core and the globule would require a better 
knowledge of the temperature profile. This is not allowed by the low 
angular resolution of the LWS observations, which averages the emission 
over a much too large region ($\sim 80\arcsec$). 
We define the total mass $\rm M_t$ of the dust condensation by 
integrating over the flux contour at $30\mjy/11\arcsec$ beam; this area
of mean size $40\arcsec$ contains a total flux $= 0.68\jy$. 
Adopting a uniform dust temperature $\rm T_d= 22\K$, we obtain
the average gas column density over the globule~: 
$2.3\times 10^{22}\cmmd$, the total mass of the globule $\rm M_t= 63\msol$,
and the average density in the globule $n(\htwo)= 4\times 10^4\cmmt$. 

Hence, we find that it is only a small mass fraction of the whole 
globule ($\sim 5\%$) which lies in the atomic surface layers, exposed to the 
strong FUV field of the stars exciting the Trifid.
The ratio of the atomic and molecular gas masses is much lower in TC2
than in the OMC~1 in Orion and in the Galactic Center, where it is about 
$16\%$. On the contrary, the ratio takes a very similar value to that in the 
young massive star-forming region W49N. This region is one of the youngest 
massive star-forming place in the galaxy, where several recently born O stars
have just started to excavate the parent cloud (Vastel et al. 2001). 
The low atomic to molecular gas ratio measured in TC2 is therefore 
consistent with a very early evolutionary age for the globule. This point is
more thoroughly addressed in Sect.~8.2. 

\section{The Molecular Condensation : zone IV}

\subsection{The Surface Layers}

The molecular content of the Bright-Rimmed Globule was first observed in the 
CO lines. The line profiles are complex and exhibit several components between 
-50 and $+50\kms$ that correspond to various physical regions in the 
molecular cloud containing the H{\small II} region (Lefloch et al. 2002). 
Because of the large extent of the CO emission in the nebula, antenna 
temperatures are a better 
approximation than main-beam brightness temperatures to the brightness 
of the CO lines. We adopted the same approximation for the $\thco$
$\jonetozero$ transition. On the contrary, we assumed main-beam
brightness temperatures for $\ceio$. 
The spatial distribution of the kinetic components  shows
that the emission of TC2 peaks at  $v_{lsr}= 7.7\kms$ (Fig.~13).   
The distribution of the emission follows closely the 
bright rim and drops abruptly beyond the ionization front. 
The antenna temperatures of the CO transitions are rather uniform 
over the globule, with values of  $\sim 30$ K for the \jtwotoone\ and 
\jonetozero\ transitions and 15 K for $\jthreetotwo$ one. 

We have tried to constrain the temperature and density distribution 
in the globule by modelling  with a radiative transfer code the excitation 
of the CO lines. A priori two zones are contributing
to the emission~: the molecular core and the PDR. 
The large ratio between the \jthreetotwo\ and the \jonetozero\ lines 
is very difficult to explain. 
Observations of the high-density gas show that the velocity shift 
between the quiescent core and the surface layers, which are accelerated by 
Radiatively-Driven Implosion, is small, of the order of $1 \kms$ 
(see Sect.~8.1). This is consistent with the symmetric 
profiles observed for all the molecular transitions (Fig.~13). 
This velocity shift is small enough that the PDR and the  dense core 
are still radiatively well coupled. Therefore, most of the radiation coming 
from the PDR on the rear side of the globule is absorbed by the dense core.
As a consequence, its contribution to the \jonetozero\ and
\jtwotoone\ lines is almost negligible and we detect mainly the emission 
of the cold core and the front side layers. We infer a kinetic 
temperature of about 30-35 K for the gas, somewhat higher than the 
dust temperature. The discrepancy could be due to the heating of the 
front side by optical photons, since the globule is immersed in ionized gas. 

The \thco\ data show widespread emission over all the globule, with typical
brightness (antenna) temperatures of $13\K$ and 12.5~K for the 
\jonetozero\ and \jtwotoone\ transitions respectively (Fig.~13). 
The \ceio\ \jonetozero\ has a main-beam brightness temperature of $2.3\K$. 
From the ratio of the $\thco / \ceio$ $\jonetozero$ brightness temperatures
(= 5.7) and assuming a standard relative abundance of 8, we derive the 
$\thco$ line opacity $\tau^{10}\simeq 0.7$. We note that the opacity
derived is fully consistent with the value of the ratio 
$\twco / \thco$ $\jonetozero$ brightness temperatures. 
An opacity $\tau^{13}= 0.7$ implies a ratio of 2 whereas the actual value
is 2.3. The mean \htwo\ density in the globule
is high enough that the $\jonetozero$ line of CO and its
isotopes are thermalized. We have used the opacity and the 
brightness of the $\thco$ $\jonetozero$ line to constrain the kinetic 
temperature and the gas column
density. The best match is obtained for a gas temperature $\rm T_k= 30\K$ 
and a column density $\rm N(\thco)\simeq 4\times 10^{16}\cmmd$.
Adopting a standard abundance $[\thco]/[\htwo]=1.6\times 10^{-6}$, we infer
the total gas column density $\rm N(\htwo)= 2.5\times 10^{22}\cmmd$
good agreement with the value obtained from the dust measurements, 
and the average gas density $n(\htwo)= 3.0\times 10^4\cmmt$, after dividing
the  gas column density by the mean globule radius.

\subsection{The dense core}

All the molecular lines observed towards the center of the globule
peak at $\rm v_{lsr}= 7.7\kms$ without any
significant variation between the tracers (Fig.~13). 
The optically thick $\hcop \jonetozero$ line has a distribution 
similar to that of CO (Fig.~14). It  was detected 
all over the globule, with
typical main-beam brightness temperatures of 2-4~K. We also observed the 
isotopic $\htcop \jonetozero$ line but failed to detect it. 

The distribution of the dense gas 
was mapped in the CS \jthreetotwo\ and \jtwotoone\ lines. 
The CS emission is less extended than $\hcop$, as 
it appears to trace the region associated with the cold dust core
(Fig.~14). 
The CS lines are bright with main-beam temperatures of the order
of 2~K, up to $\rm \sim 4~K$ (4.3~K and 4.0~K for the $\jtwotoone$ and the 
$\jthreetotwo$ lines respectively) at the position $(72\arcsec,-135\arcsec)$.  
The CS~$\jtwotoone$ transition traces a region approximately circular
with a typical size of $5\times 10^{17}\cm$ (HPFW), marginally
resolved by the telescope beam.  
The higher angular resolution of CS \jthreetotwo\ data 
reveals two gas components~: a weak, extended, component which overlaps 
very well with the globule, as traced in $\twco$ and the millimeter
continuum, and a strongly peaked condensation in the center of the globule 
(Fig.~14)~: the central CS core. The core is slightly elongated in the
North-South direction. Its dimensions (beam-deconvolved) are 
$5.2\times 10^{17}\cm$ by $6.7\times 10^{17}\cm$, very similar to the 
size of the the CS~$\jtwotoone$ emitting region. Hence both lines are
probing the same region. 
As can be seen in Fig.~15, there is no evidence of a velocity shift in the 
CS~$\jtwotoone$ emission peak along the major axis. The same results holds 
for the 
the \jthreetotwo\ transition and the $\hcop$ line. This indicates that 
the molecular gas of the globule, in particular the dense core,  is in a 
quiescent stage.

In order to determine the physical conditions in the dense core, we carried
out an LVG analysis on the \jtwotoone\ and \jthreetotwo\ lines assuming 
a gas kinetic temperature of $20\K$, suggested by the millimeter continuum 
observations. 
Three positions were studied~: the center of the core and two other 
positions offset by 
$15\arcsec$. The velocity width was estimated 
from a gaussian fit to the line profiles~: $\Delta v= 1.3\kms$.
In the central position we find a density 
$n(\htwo)= 2.0\times 10^5\cmmt$ and a column density 
$\rm N(CS)= 1.8\times 10^{13}\cmmd$. The line opacities are relatively
low, with $\tau^{21}= 0.42$ and $\tau^{32}=0.70$. 
Similar densities are obtained around the center ($\sim 2\times 10^5\cmmt$)
and at the brightness peak, $15\arcsec$ South of the center. At this position, 
we find $\rm N(CS)= 3.0\times 10^{13}\cmmd$.
Hence, from the \htwo\ core density and the beam size, we find
a core mass $\sim 28\msol$, in good agreement with the dust estimate.
Adopting a typical abundance $\rm [CS]/[\htwo]= 5\times 10^{-10}$, we 
estimate the size of the CS emitting region $\ell= 3.0\times 10^{17}\cm$. 
This value is consistent with the the size HPFW derived from the 
maps of velocity-integrated emission at 3 and 2mm.

The picture coming out of the molecular line and thermal dust continuum 
analysis is that the globule
consists of a cold central core of density $\rm n(\htwo)= 2-3\times 10^5\cmmt$
surrounded by an envelope slightly warmer in the outer layers, at a
temperature $\rm T\sim 30\K$, and density of about $3\times 10^4\cmmt$. 

\section{Radiatively-Driven Implosion of TC2}

We have recalled briefly in the Introduction the overall evolution of 
a photoionized globule. The time evolution of a photoionized globule, the
density and the velocity fields, were studied numerically and presented 
in LL94 (see in particular their Fig.~4) under a wide range of 
ionization conditions, including those typical of bright-rimmed 
globules. The evolution of 
the photoionized globule is determined by the two following parameters~: \\ 
a)~$\gamma= \eta \alpha n_e R_g/c_i$ ~: the ratio of the impinging photons 
consumed to balance recombinations to those used to ionize neutral material.
For TC2, we obtain $\gamma= 15$. Hence,  most of the
Ly-c photons are consumed in the ionized gas layer surrounding the neutral 
condensation (see also Sect.~4.1). \\
b)~$\delta= n_e/n(H)$~: the ratio of the ionized gas density with
respect to the neutral gas density.  This factor is directly related 
to the overpressure exerted by the ionized gas, and the intensity 
of the shock driven in the condensation.
The mean globule density, ahead of the PDR, is the most difficult term 
to estimate. It is probably a good approximation to assume a density
comprised between the PDR density ($ n(\htwo)= 10^4\cmmt$) and the molecular
gas traced by $\thco$ ($n(\htwo)= 3\times 10^4\cmmt$).
Hence, we estimate that $\delta$ lies in the range $0.03 - 0.09$. 
The values of $\delta$ and $\gamma$ found for TC2 are typical of 
a bright-rimmed globule. Following the convention defined by LL94, 
this corresponds to region IV in the $(\delta - \gamma)$ plane  (see their 
Fig.~3). In this case, the whole evolution is governed 
by the propagation a D-critical I-front preceded by a shock front. In their
study, LL94 showed that  
the morphology, the density and velocity structure of a photoionized 
globule mainly depend on the duration of the ionization, i.e.
the time elapsed since the illumination began. It does not depend
critically on the ``real'' values of $\delta$ and $\gamma$. 

A direct estimate of the total (ram + thermal) external pressure at the 
surface of the cloud yields~:
$P_i/k_B= 2.11 \times 2 n_e T_e =  4.2\times 10^{7}\K\cmmt$. 
We can estimate the inner pressure from the kinematic motions as measured  
by the the linewidth of the \thco\ and CS transitions~: $\Delta v= 1.4\kms$.
We find  a kinetic pressure $P_{k}/k_B= 2.3\times 10^{6}\K\cmmt$. 
This is 20 times weaker than the outer pressure. 
Therefore, the inner pressure cannot sustain the globule against the
overpressure of the ionized material. 
LL94 showed that in the course of its evolution, the pressure of the
ionized gas increases as the radius of the globule decreases
(Sect.~5.1 in LL94). In other words, the overpressure at the surface of the
globule was less in the past. 

\subsection{Observational Evidence}

Figure~15 shows the emission of the dense molecular gas 
in the $\hcop \jonetozero$ and CS~$\jtwotoone$ lines along a cut in 
declination across the border of the globule.
The $\hcop \jonetozero$ line is optically thick and is therefore 
well suited to trace the motions in the surface layers of the globule.
Next to the $\rm HCO^{+}$ emission main peak at $7.7\kms$,
we detect a second component at ``blue'' velocities shifted 
by $0.7\kms$. 

This feature appears as bright as the main body gas emission. 
It is unambiguously detected in the CS ~$\jtwotoone$ line. 
It is detected only  along the border of the globule, 
$15\arcsec$ North with respect to the center, and {\em not inside} the main 
body. It disappears again $30\arcsec$ North of the 
center, at the tip of the globule.
The weakness of the line at these declinations is due to the 
small beam filling factor
as the major fraction of the beam solid angle points towards the 
ionized gas in the \Hp\ region. 
The low SNR of the spectrum prevents from leading any quantitative
analysis. 

This component is not related with prostostellar activity. 
First, the blueshifted component is detected {\em only} at the border of the 
globule. There is no evidence of a blueshifted component 
towards the  dust emission peak $15\arcsec$ South (more than 
$2\times 10^4$~AU away from the border of the globule) where 
the protostar is expected to be found (see Sect.~8.2).
Second, the blueshifted component is kinematically
separated from the main body gas emission and it does not exhibit the 
typical ``wing'' profile of molecular outflows. 
Third, there is no evidence for an additional, redshifted, component which 
would trace the other wing of the molecular outflow. 

On the contrary, the profile of this secondary kinematic component is much 
more suggestive of a shock  propagating into the globule from the 
{\em rear} side. The detection of this secondary kinematic component 
at the border of the globule, well separated from the main body gas 
emission, is typical of a globule in the early collapse phase 
and testifies that a shock
is propagating into the globule from the surface. This is in
agreement with the mid-IR analysis which showed that the globule is
photoionized on the rear side. {\em This secondary molecular component 
is the kinematical signature of the PDR}.
It is completely unresolved by the 30m telescope.
However, since the far-infrared observations indicate a typical size of 
$10^{17}\cm$ for the PDR ($\sim 4\arcsec$ at the distance of the Trifid), 
it could be resolved out by millimeter interferometers. 

\subsection{Comparison with models}

We compare the physical properties of TC2 with the numerical modelling 
of a photoionized globule presented in LL94. The simulation
was done for ionization parameters similar to TC2~: 
$\Delta= 0.1$ and $\Gamma= 10$
($\Delta$ and $\Gamma$ are the initial values of the parameters $\delta$ and 
$\gamma$). 
It is therefore possible to relate the properties of the {\em observed} 
globule to the simulated cloud via the simple scaling (LL94)~:
$r\rightarrow kr$, $t\rightarrow kt$, $\rho\rightarrow k^{-1}\rho$.  

As mentioned above, TC2 is undergoing the collapse phase, when a shock
front propagates into the neutral gas. Based  on a morphological comparison
with the simulated globule (Figure~4 in LL94), we find the best 
match for an age comprised between $0.13\Myr$ and $0.18\Myr$ in
the computation. 
At $t= 0.13\Myr$ (Fig.~4b), a dense core has formed below the surface of the 
globule, which still exhibits a ``barnacle'' shape. Later on, at $0.18\Myr$,
(Fig.~4c), the bulk of the material has collapsed onto the main axis. 
The globule has now adopted an elongated shape. 
The best morphological match is obtained for an intermediate time 
$\tau \simeq 0.15\Myr$. The numerical modelling indicates that the 
amplitude of the
secondary kinematic component at such early photoionization stages 
is typically $1-2\kms$ (see their Fig.~14a). This velocity shift could be 
smaller depending on the inclination angle of the globule surface with 
respect to the line of sight. The $0.7\kms$ difference observed in TC2 
between both kinematic components is therefore fully compatible with the 
numerical results. 

At that stage, the simulated globule has lost about $25\%$ of its initial 
mass (see Fig.~8  in LL94).
Applying this result to TC2, it means that the initial mass of the globule 
was $\approx 80\msol$. Since the initial mass of the {\em simulated} globule  
is $20\msol$, we can derive the scaling factor relating the
properties of the observed globule to the model~: $k\simeq 2$. 
The radius of the {\em scaled} (simulated) globule is then 
$\sim 6\times 10^{17}$~cm whereas we measured $4.0\times 10^{17}$~cm. We 
note however that the density is assumed to be uniform in the simulated 
globule, whereas a core-envelope density structure, as in TC2, would lead 
to a smaller radius for the same ionization conditions. 
The duration of the photoionization, scaled from the simulations, is 
$0.3\Myr$. This compares very well with the kinematical age of 
the Trifid, as measured from the luminosity of the exciting
star and the expansion of the nebula (Lefloch \& Cernicharo, 2000). It is 
therefore very likely that the 
globule was very early exposed to the ionizing radiation, as soon as
the O star turned on.  

\section{Star Formation}

\subsection{The HH~399 jet}

The first evidences of ongoing star formation in TC2 were presented 
by CL98, based on a $\sulfp$ image of the HH399 jet running out of the 
bright rim of TC2. 
The jet has been studied in the optical by Rosado et al. (1999) and 
CL98. These observations revealed the fragmented structure of the jet 
and could identify several ``knots'' from the head to the base of the jet, 
A-G, A being the most remote knot (see Fig.~5b). 

Our VLA observations detect only the external part of the HH~399 jet. 
We could detect three ``clumps'' which coincide with the knots 
dubbed A, C and D by Rosado et al. (1999). 
The clumps are unresolved in the transverse direction by our observations. 
The flux peaks are rather similar (between 70 and 
$90\mujy\, \rm beam^{-1}$). We take $0.85\arcsec$ 
($\approx 2\times 10^{16}\cm$) as an 
upper limit to the diameter of the jet, and adopting a temperature between 
5000 and $10^4\K$ we derive average electronic densities of 
$1.0-1.5\times 10^3\cmmt$ in the clumps.  
Our estimate is in agreement with Rosado et al. (1999) for knot~A. 
There are larger discrepancies for knots  C and D, which the authors
estimate to have densities of $3\times 10^3\cmmt$ and 
$8\times 10^3\cmmt$ respectively. This could be due to clumpiness 
in the jet structure. 
This can be also explained by our assumption
on the geometry of the jet, apart from the uncertainties in the 
3.6cm flux (the noise is $\sim 15-20\mujy\, \rm beam^{-1}$). 
For instance, assuming a jet diameter  ten times less would yield an
electron density of $ 3\times 10^3\cmmt$, closer to the estimates
derived at optical wavelengths. 

As soon as the jet runs out of the PDR, it is exposed to the ionizing
radiation of the exciting star. 
Integrating over the contour at $20\mujy\, \rm beam^{-1}$,
we obtain a total flux $S_j= 0.36\mjy$ for the jet, and an average 
emissivity $\rm EM= 8.8\times 10^3\pc\,cm^{-6}$. The recombination
rate per surface unit of the jet is $7.3\times 10^9\cmmd\smu$. 
This is only half the ionizing flux impinging on the globule, as 
estimated in Sect.~3.1. Hence, there are enough Ly-c photons to maintain 
the jet fully ionized once it propagates out of the globule. 
HH399 is one of the finest examples of photoionized jets reported so 
far (see also Reipurth et al. 1998). 

We searched for some molecular outflow emission  which would be the
counterpart to the optical jet but did not find any evidence of such a 
phenomenon. This is probably because of the 
unfavourable orientation of the jet, very close to the plane of the sky 
(Rosado et al. 1999 estimate an inclination lower than $4\deg$). 
We also searched for SiO emission as it is a tracer of young 
protostellar outflows, associated with Class 0 and/or massive 
sources (Bachiller, 1996; Lefloch et al. 1998). 
Again, this search yielded only negative results. This suggests that the 
powering
source has probably reached the Class~I stage, characterized by a less 
active accretion phase than the Class 0, and has a typical evolutionary age of 
a few $10^5\yr$. If this age is correct, it means that star formation began
approximatly at the time when the ionizing star HD~164492A turned on and 
ignited the nebula. 

\subsection{The protostellar source}

The observation of the molecular outflow emission and the overlap
between the wings usually provides a good determination of the position 
of the driving source. Since this method cannot be applied here, in the
absence of molecular outflow, the only indications on
the source location are provided by the VLA observations. 
We have detected two small components inside the molecular core of TC2
(Fig.~16). Remarkably, they are both aligned with the 
jet propagation axis. Both components  are unresolved and are only 
marginally detected with fluxes of 50 and $70\mujy\, \rm beam^{-1}$ 
respectively at a level of $3\sigma$ and $4.5\sigma$ respectively. 
It is not clear whether these components trace some material associated 
with the counterjet or some material in the protostellar
environment of the driving source itself. However, since they are located 
close to the millimeter dust emission peak we favor the latter hypothesis. 
In the mid-infrared, as observed with ISOCAM in the  broad band filters, 
there is no evidence of point-source emission in the globule. This indicates 
that the powering source of HH399 is a much less luminous and massive
object than the protostar detected in 
Southwestern molecular cloud (Lefloch \& Cernicharo 2000).  
A close inspection of the continuum emission between 13 and 
$\rm 15\mu m$ in the globule however reveals a 
local maximum of compact, unresolved, emission
at $\Delta\alpha= +70\arcsec$, $\Delta\delta= -124\arcsec$. This source is 
spatially separated from the PDR, though 
close to it ($\approx 7\arcsec$). It coincides with the VLA sources
to better than $2\arcsec$, actually falling between them. 
Altogether, these observations give strong
support for a physical association between the VLA component(s) and the
driving source of HH399 jet. 

The bolometric luminosity of the protostellar source is difficult to 
estimate since the PDR contributes to the luminosity of the globule too. 
We estimate the protostellar luminosity by taking into account 
only the cold dust component estimated from the fit of the SED~: 
$\rm L\sim 520\lsol$.
An upper limit is obtained by integrating under the spectral energy
distribution~: $\rm L_{max}= 1200\lsol$. 
This value is typical of intermediate-mass protostars but we stress again
that the actual luminosity could be lower. 

\subsection{The ultimate fate}

The radiation field has a deep impact on the Herbig-Haro jet as soon as it
escapes the globule. However, it is less clear how much the star forming
process itself is perturbed. In the present stage, the size of the PDR 
represents approximately one third of the globule's radius. It is the 
low-density envelope, i.e. the
mass reservoir, which is affected by the FUV field, and not the dense core 
where the material is being accreted.  
An important paramater in the evolution of the globule is its mass-loss
rate, which is determined by the ionizing flux and the radius of the
globule (see LL94)~: 

\begin{equation}
\dot{M} = 14 \left(\frac{\Phi_i}{10^7\cmmd\smu}\right)^{1/2}
\left(\frac{R_g}{1\pc}\right)^{3/2} \msol\Mymu\nonumber 
\end{equation}

For TC2, we estimate a present mass-loss rate of $34\msol\Mymu$. From 
comparison with numerical simulations, we found that the globule lost about 
$20\msol$ since the ionization began. 
The globule has evaporated half the mass of its envelope until now and its
expected lifetime is therefore $\tau_g= M/\dot{M}\simeq 1.9\Myr$. This is 
a lower estimate as the mass-loss rate increases with radius.  
On the other hand the active accretion phase is known to last a few
$10^5\yr$ (Andr\'e et al. 2000). 
Therefore, the life expectancy of the globule is long enough to allow 
the protostellar objects to complete smoothly the accretion phase 
and turn into stars before the globule wholly evaporates. 
TC2 is going to develop progressively a low density tail, turning into 
a so-called Cometary Globule.
We speculate that within a few $10^5\yr$ the situation should be very
similar to the gas fingers observed in the Eagle Nebula (White et
al. 1999), where
the strong UV field is evaporating the ultimate dust and gas layers of the 
stellar nests, unveiling the star(s) formed in the early stages of the
nebula. 

\section{Conclusions}

We have carried out a multiwavelength study of bright-rimmed globule TC2
in the Trifid nebula. 
The globule lies almost in the plane of the sky, immersed in  the 
low-density gas of the $\Hp$ region ($\rm n_e\sim 50-100\cmmt$). 
It is illuminated by the O star HD~164492A, mainly on the rear side. 
The globule consists of a very dense core of cold 
gas and dust (T= 22~K, $\rm n(\htwo)= 3\times 10^5\cmmt$), of small dimensions 
surrounded by a lower-density envelope 
($\rm n(\htwo)= 3\times 10^4\cmmt$). Its mass ($\simeq 63\msol$) is 
typical of bright-rimmed globules, equally distributed between the core 
and the envelope (27 and $36\msol$ respectively). 
The ionization conditions at the surface of the globule were  
determined from VLA and ISO (LWS and SWS) observations. The impinging 
ionizing flux is $1.4\times 10^{10}\cmmd\smu$; it creates an ionization 
front and a photoevaporated envelope of density $1-2\times 10^3\cmmt$ 
at the surface of the globule. 

The PDR is traced by the emission of the PAHs bands at 6.2, 7.7, 8.6 
and $\rm 11.3\mu m$.
The observed variations of the relative intensities can be accounted
for by the change in the excitation conditions. 
We find that the intensity of the $\rm 11.3\mu m$ band drops outside of the 
PDR in the photoionized layers, 
as a consequence of the ionization of the PAHs.
The $\rm 7.7\mu m$ band is still detected in the photoionized
envelope of the globule, though at a weaker level. 
Despite the  high visual extinction of the globule, the 
$\rm 7.7\mu m$ PAH band, excited in the PDR on the rear side is detected 
in the body of the
globule thanks to a minimum in the absorption of the ices and the dust 
in this wavelength range. 

Millimeter line observations reveal the kinematical signature of the PDR 
which precedes the ionization front as a shock moving into the globule. 
The relative projected velocity is weak,  $\approx 0.7\kms$.
Comparison with models of photoionized globules (LL94) indicates that TC2 
has been  exposed to the ionizing radiation for $3\times 10^5\yr$, 
almost as soon as the exciting star of the nebula turned on. 
The structure of the photon-dominated layer has been derived from the 
ISO far-infrared line and continuum observations. The agreement between 
all the data and the model of K99 is good. The intensity of the radiation 
field at the
surface is $\rm G_0\simeq 1000$. The PDR has a typical column density of 
$2\times 10^{21}\cmmd$ and a molecular hydrogen density  
$\simeq 10^4\cmmt$. The FUV field heats the dust to a temperature 
$\simeq 46\K$ whereas the average gas temperature is found close to 
$300\K$. The emission of the $\Sip$~$\rm 34.8\mu m$ line in the PDR 
can be accounted for by assuming a silicon gas phase abundance 
of $6\times 10^{-6}$, i.e. $\simeq 17\%$ the solar value. 

The globule is currently undergoing star formation. Our observations
show that the globule is forming a low- or intermediate-mass  
star of luminosity $\leq 500 \lsol$. The protostar powers a Herbig-Haro
jet, which appears fully ionized outside of the globule. 
No molecular counterpart (outflow) to the optical jet has been found. In 
particular, no SiO 
emission was found. This implies that the source is already in an
intermediate stage between Class 0 and Class I, or even a full Class I
member. The evolutionary age of the source is therefore typically 
a few $10^5\yr$, which suggests that star formation in TC2 probably 
started in the large burst which accompanied the 
birth of HD~164492A. As a result of the photoevaporation of the surface
layers, the globule has evacuated about half the mass in its envelope until
now but we have not found any evidence that the birth process itself has
been perturbed. On the contrary, 
the photoevaporation rate is low enough to leave ample time for protostars 
to reach safely the ultimate stages of star formation. 

\acknowledgements

We acknowledge Spanish DGES for this supporting research 
under grants PB96-0883 and ESP98-1351E. We thank J.R. Pardo for help with 
the observations at the CSO. LFR is grateful
to CONACyT, Mexico, for its support.
This research made use of SIMBAD.

\clearpage

{\small\noindent Fig.~1  -- 
Schematic drawing of the TC2 region~:
(I)~the H{\small II} region between the exciting star (HD~164492A)
and the bright-rimmed globule; (II)~the ionization front (IF) and the 
expanding 
envelope of photoionized gas around the globule; (III)~the Photon-Dominated 
Region (PDR) and the shock ahead of the IF; (IV)~the molecular core of TC2. 
}
\smallskip
\label{fig1}
\smallskip                                                                

{\small\noindent Fig.~2 -- 
[$\rm H\alpha$] image of the Trifid Nebula observed with 
the NOT 
telescope (CL98). The location of Bright-Rimmed Globule TC2 is marked by a 
white 
rectangle. The positions observed with ISO/LWS and the telescope beams
are indicated by circles. 
}
\smallskip
\label{fig2}

{\small\noindent Fig.~3  -- 
Thermal dust emission of TC2 observed at 1.25mm with the MPIfR 19 channel
bolometer array (thin white contours) superposed on the optical $\rm \sulfp$ 
emission of the region. Contours range 
from 10\% to 90\% of the peak flux ($160\mjy /11\arcsec$ beam). 
In thick white contours,  we have superposed the mid-IR emission, as 
observed with 
the ISOCAM LW10 filter ($\rm 8-15\mu m$). First contour and interval are 
160 and 20 MJy/sr 
respectively.  
In thin black contours is superposed the 3.6cm free-free emission observed 
at the VLA. Contours are 0.4, 0.8, 1.1, 1.5, 1.9 mJy/beam. 
}

\smallskip
\label{fig2}

{\small\noindent Fig.~4 --  
Magnified View of TC2 as observed in the 
[S{\small II}] lines $6717, 6731\AA$. 
The white circle marks the TC2 region encompassed with the LWS beam.  
The rectangles mark the regions observed with ISO/SWS, and delineate the
aperture of the detector (band 4). The center of 
the SWS fields are marked by stars. 
}
\smallskip
\label{fig3}

{\small\noindent Fig.~5 -- 
(a)~Contour map of the free-free emission at 20~cm
obtained with the
VLA (CL98), superposed on the optical [S{\small II}] image of the globule.
First contour and contour interval are $1.5\times 10^{-2}$ and
$2.5\times 10^{-3}$~Jy/beam respectively. \\
b)~Contour map of the free-free emission at 
3.6~cm obtained with the VLA,  superposed on an
 [S{\small II}] image of the head of bright-rimmed globule TC2 (greyscale).
The first contour and contour interval are $4\times 10^{-5}$ and 
$2\times  10^{-5}$~Jy/beam respectively.
}
\smallskip
\label{fig5}

{\small\noindent Fig.~6 -- 
Fine structure atomic and ionic lines detected at the three positions 
observed with ISO/SWS and at the two positions observed with ISO/LWS.
}
\smallskip
\label{fig7}

{\small\noindent Fig.~7 -- 
Mid-infrared emission as detected 
with the CVF in the PAH bands at 6.2, 7.7, 8.6 and $\rm 11.3\mu m$
(greyscale and contours), in the  $\Nep$~$\rm 12.7\mum$ (with the PAH 
$12.7\mu m$ band) and the $\Nepp$~$\rm 15.5\mum$ line and in the 
continuum between 13 and $\rm 15\mu m$ (greyscale). 
We have superposed (contours) the 
PAH~$\rm 11.3\mu m$ band emission on the panels of the 
$\rm 12.7\mum$, $\Nepp$, and $13-15\mu m$ emissions. 
The location of the spectra shown in Fig.~9 is marked with white squares. 
}
\label{fig6}

{\small\noindent Fig.~8 -- CVF Emission in the range $\rm 5-16 \mu m$ 
at a few positions towards TC2. The locations are marked in the lower 
right panel of Fig.~2d. 
{\em (a)} and {\em (b)}~: photoionized region (positions A and B respectively).
{\em (c)} and {\em (d)}~: photon-dominated region  (at C and D resp.); 
{\em (e)}~: reference position (position E); {\em (f)}~: core of TC2
(position F). 
In panels {\em a)-d)}, we have superposed (thin contour) the spectrum of the 
reference position scaled  to match the intensity of the 
$\rm 7.7\mu m$ band. 
}
\label{fig8}
\smallskip

\smallskip
{\small\noindent Fig.~9  -- Comparison of the mid-infrared emission in the
  core of the globule (position F, bottom spectrum) with the emission in the 
PDR at position D without absorption (top spectrum) and with an absorption 
of 20 mag (dashed). The flux scale of the spectra in the core of TC2 and in
the absorbed PDR has been divided by 2.0 
}
\label{fig9}
\smallskip                                                                

{\small\noindent Fig.~10 --  
Large-Velocity Gradient analysis of the $\Oi$ lines observed with the LWS.
Isoflux contours in the temperature-density diagram
for the $\Oi$~$\rm 63\mu m$ (solid) and $\rm 145\mu m$ (dashed). for three
oxygen column densities. For each line, we have also indicated the upper
and lower limits at $20\%$.  
}
\smallskip
\label{fig10}

\smallskip
{\small\noindent Fig.~11 -- Mid-Infrared emission 
observed  between 2 and $\rm 45\mu m$ with the
Short Wavelength Spectrometer onboard ISO, towards 3 positions
centered respectively on the PDR of TC2 
at the offset position 
$\Delta\alpha= +71\arcsec \quad \Delta\delta= -120\arcsec$ 
(central panel), the main body of the globule 
(lower panel) and the bright rim and the \Hp\ region (upper panel) 
The thin line draws the fit to the SED.
}
\label{fig11}
\smallskip

{\small\noindent Fig.~12  -- 
Spectral Energy Distribution of TC2 after subtracting the extended emission
of the nebula. We have
indicated (square) the $1250\mum$ flux (integrated over the ISO/LWS beam). 
We show in dashed the fits to the warm and cold dust component. The fit to
the whole emission is indicated by the solid line. 
}
\smallskip
\label{fig12}
\smallskip

{\small\noindent Fig.~13 -- 
Montage of the molecular lines detected in the core of TC2.
The line fluxes are given in the antenna temperature scale. 
The dashed line indicates the systemic velocity of the dense core 
$\rm v_{lsr}= 7.7\kms$. 
}
\label{fig13}
\smallskip                                                                

{\small\noindent Fig.~14 -- 
Contour maps of the velocity-integrated emission  
$\hcop \jonetozero$, CS~$\jtwotoone$, $\twco \jonetozero$,  
$\twco \jtwotoone$,
CS~$\jthreetotwo$ and $\twco \jtwotoone$. The emission was integrated
between 5 and $10\kms$. Contours range 0.1, 0.2... to 0.9 times the 
emission peak. The contours are superposed on a 1.3mm continuum emission
map of the globule. 
}
\smallskip
\label{fig14}
\smallskip

{\small\noindent Fig.~15 -- 
Molecular gas emission in the CS~$\jtwotoone$ and $\hcop \jonetozero$ lines 
in a cut across the Photon-Dominated Region in North-South direction. 
The dashed lines indicate the systemic velocity of the main body gas in TC2
($\rm v_{lsr}= 7.7\kms$) and of the second kinematic component detected 
in the PDR ($\rm v_{lsr}= 7.0 \kms$). 
}
\smallskip
\label{fig15}
\smallskip

{\small\noindent Fig.~16 -- 
VLA emission at 3.6cm (contours) superposed on the continuum emission
between 13 and $\rm 15\mu m$ as observed with the CVF. 
}
\smallskip
\label{fig16}
\clearpage
\pagebreak 

\begin{table*}
\begin{center}
\caption[]{List of Observations. 
} 
\label{lines}
\begin{tabular}{llrcr}\\
\tableline\tableline
Zone    & Instrument & Observations & Wavelength & Figures  \\
\tableline

I ($\Hp$ region), & VLA & Free-free radiation & 3.6 cm, 20 cm & 5a, 5b, 16 \\
II (Bright Rim)   & NOT & $\rm H\alpha$, $\sulfp$ & Optical    & 2, 4\\
   & SWS, CVF, LWS & Atomic Lines~: &  $\rm 2.5-197\mu m$ & 6,7 \\    
   &         &   $\Nep$, $\Nepp$, $\sulfpp$, $\Npp$, & & \\
   &         &    $\Opp$, $\Sip$, $\Cp$             & & \\
           & SWS      & Continuum (dust) &  $\rm 2.5-45\mu m$   & 11 \\
           &                &                &                  & \\
III (PDR)  & ISOCAM         & LW10           & $\rm 8-15\mu m$ & 3 \\
           &    CVF         & PAH bands      & $\rm 5-17\mu m$ & 7,8 \\
    &  SWS, LWS      & Continuum (dust) & $\rm 2.5-197\mu m$ & 11, 12 \\ 
     &   SWS, LWS  &  Atomic  Lines~:  &  $\rm 2.5-197\mu m$ & 6,7\\
           &                 & $\Oi$, $\Cp$, $\Sip$   &         & \\
           &                 &                &                     & \\ 
IV (Molecular Core) & LWS &  Continuum (dust) & $\rm 45-197\mu m$ & 12 \\
                    & IRAM-30m & Continuum (dust) &1.25mm        & 3 \\
                    &          & CO, $\thco$, $\ceio$, & 1-3mm  & 13, 14, 15 \\
                    &          & $\hcop$, SiO, CS       &       & \\
                    & CSO      & CO                     & 0.8mm  & 13 \\
                 
\tableline
\end{tabular}
\end{center}
\end{table*}

\begin{table}
\begin{center}
\caption[]{Millimeter lines observed towards TC2~: frequency, 
telescope beamwidth and efficiency. 
} 
\label{lines}
\begin{tabular}{llrcr}\\
\tableline\tableline
    & Line & Frequency & Beamwidth & $\rm B_{eff}$  \\
     &  & GHz      &$\arcsec$  &                \\
\tableline
$\htcop$ & $\jonetozero$    & 86.75429  & 28 & 0.77 \\
SiO     & $\jtwotoone$     & 86.84700  & 28 & 0.77 \\
$\hcop$ & $\jonetozero$    & 89.18852  & 27 & 0.75 \\
CS      & $\jtwotoone$     & 97.98097  & 24 & 0.71\\
$\ceio$ & $\jonetozero$    & 109.78218 & 22 & 0.68 \\
$\thco$ & $\jonetozero$    & 110.20135 & 22 & 0.68 \\
CO      & $\jonetozero$    & 115.27120 & 21 & 0.67 \\
SiO     & $\jthreetotwo$   & 130.26870 & 18 & 0.58 \\
CS      & $\jthreetotwo$   & 146.96905 & 16 & 0.53\\
SiO     & $\jfivetofour$   & 217.10494 & 11 & 0.42 \\
CO      & $\jtwotoone$     & 230.53800 & 10 & 0.39 \\
CO      & $\jthreetotwo$   & 345.79599 & 22 & 0.75 \\
\tableline
\end{tabular}
\end{center}
\end{table}

\begin{table}
\begin{center}
\caption[]{Fine structure lines detected with the SWS. The fluxes are 
uncorrected for the mid-IR extinction. 
\label{flux_sws}}
\begin{tabular}{lcccrrc}\\
\tableline\tableline
Line & Band & $\lambda $ & Aperture & $\rm F_{\nu}^{(0,+0)}$ &  
$\rm F_{\nu}^{(0,+20)}$ & $\rm F_{\nu}^{(0,-20)}$ \\
       &  & $(\mu m)$ & $(\arcsec \times \arcsec)$ & $\rm W\cmmd$ & 
$\rm W\cmmd$ & $\rm W\cmmd$ \\
\tableline
$\Nep$     & 3A & 12.8 & $14\times 27$ & $(7.5\pm 0.7) (-19)$ & 
$(9.0\pm 0.9) (-19)$ & ($8.4\pm 0.8) (-18)$ \\
$\Nepp$($^{a}$)    & 3A & 15.5 & $14\times 27$ & $(6.9\pm 0.7) (-19)$ & 
$(9.0\pm 0.9) (-19)$ & \_ \\
$\sulfpp$  & 3C & 18.7 & $14\times 27$ & $(1.3\pm 0.1) (-18)$ & 
$(1.3\pm 0.1) (-18)$ & $(9.0\pm 0.9) (-19)$\\ 
$\sulfpp$  & 4  & 33.5 & $20\times 33$ & $(4.2\pm 1.2) (-18)$ & 
$(3.9\pm 1.0) (-18)$ & $(3.6\pm 1.0) (-18)$\\    
$\Sip$     & 4  & 34.8 & $20\times 33$ & $(9.6\pm 2.9) (-19)$ & 
$(9.6\pm 2.9) (-19)$ & $(8.3\pm 2.5) (-19)$ \\
\tableline
\end{tabular}
($^{a}$)~Unreliable due to large statistical errors.
\end{center}
\end{table}

\begin{table}
\begin{center}
\caption{Fine structure atomic lines detected with the LWS. 
\label{flux_lws}
}
\begin{tabular}{lrccccc} \\
\tableline 
\tableline
Line  & Band  & Beam Size & $\lambda $ & $\rm F_{\nu}^{On}$ &  
$\rm F_{\nu}^{Off}$ & TC2 (``On'' - ``Off'') \\
       &    & (\arcsec)   & $(\mu m)$ & $\rm W\cmmd$ & $\rm W\cmmd$ 
& $\rm W\cmmd$ \\
\tableline
$\Opp$  & SW2 & 84 & 51.7  & $(7.1\pm 1.4) (-18)$ & ($2.5\pm 0.5) (-18)$ & 
4.6 (-18) \\
$\Npp$   & SW2 & 84 & 57.2  & $(4.9\pm 1.0) (-18)$ & ($2.2\pm 0.4) (-18)$ & 
2.7 (-18) \\
$\Oi$   & SW3 & 86 & 63.3  & $(9.8\pm 2.0) (-18)$ & $(2.6\pm 0.5) (-18)$ & 
7.2 (-18) \\
$\Opp$  & SW5 & 80 & 88.4  & $(9.0\pm 1.8) (-18)$ & $(4.0\pm 0.8) (-18)$ & 
5.0 (-18) \\
$\Np$  & LW2  & 68 & 122.0 & $(1.4\pm 0.3) (-18)$ & $(7.7\pm 1.5) (-19)$ & 
6.3 (-19) \\
$\Oi$  & LW3  & 70 & 145.5 & $(5.8\pm 1.2) (-19)$ & $(1.5\pm 0.3)
(-19)^{*}$ &  4.3 (-19) \\
$\Cp$  & LW4  & 68 & 157.8 & $(9.8\pm 2.0) (-18)$ & $(7.1\pm 1.5) (-18)$ & 
2.5 (-18) \\
\tableline
\end{tabular}
\end{center}
\end{table}

\begin{table*}
\begin{center}
\caption[]{Physical Properties of TC2. \label{flow}}
\begin{tabular}{ll}\\
\tableline
\tableline
FUV Radiation Field ($\rm G_0$)         & 1000 \\
Infrared Luminosity                     & $1200\lsol$ \\     
Total Mass ($^{a}$)                     & $63\msol$ \\ 
Mean Radius $\rm R_g$                   & $4\times 10^{17}\cm$  \\
Mean $\htwo$ Column Density             & $2.3\times 10^{22}\cmmd$ \\
                                        &  \\
Core                                    & \\   \hline\noalign{\smallskip}
Dimensions ($10^{17}$~cm)               & $3\times 8$  \\
Dust Temperature                        & 22~K \\
Maximum $\htwo$ Column density ($^{a}$) & $8.0\times 10^{22}\cmmd$ \\
Mean $\htwo$ Column density ($^{a}$)    & $5.1\times10^{22}\cmmd$ \\
Mass                                    & $27\msol$   \\
$\htwo$ Density                         & $3\times 10^5\cmmt$ \\
Central source luminosity               & $\leq 500\lsol$ \\ 
                                      & \\      
Molecular Envelope                    & \\ \hline\noalign{\smallskip}        
$\htwo$ Density ($\thco$)             & $3.0\times 10^4\cmmt$ \\
Gas Temperature                       & $30\K$ \\
                                      & \\  
Photon-Dominated Region               &  \\ \hline\noalign{\smallskip}        
Thickness                             & $1\times 10^{17}\cm$ \\
Dust Temperature                      & 46~K   \\
Atomic Gas Temperature (OI)           & 300~K \\
Gas Column Density ($^{a}$)           & $2.0\times 10^{21}\cmmd$ \\
Oxygen Column Density                 & $7\times 10^{17}\cmmd$ \\
Oxygen Abundance                      & $3.0\times 10^{-4}$ \\
Carbon  Column Density                & $3\times 10^{17}\cmmd$ \\
Carbon Abundance                      & $1.4\times 10^{-4}$ \\
Density ($\htwo$)                     & $1.0-6.0\times 10^4\cmmt$ \\           
Atomic Gas Mass (CII)                 & $3-4\msol$ \\
Velocity field                        & Radiatively-Driven Implosion \\
                                      &        \\
Photoionized Envelope (Bright Rim)    &  \\ \hline\noalign{\smallskip} 
Thickness                             & $\sim 5\times10^{16}\cm$\\
Electron Density                      & $1.0-2.0\times 10^3\cmmt$\\
\tableline
\end{tabular}
\tablenotetext{a}{from dust observations.}
\end{center}
\end{table*}

\end{document}